\documentclass[aps,prd,preprint,groupedaddress]{revtex4-2}

\usepackage{amssymb}
\usepackage{amsmath}
\usepackage[utf8]{inputenc}
\usepackage{mathtools}
\usepackage{graphicx}
\usepackage{subcaption}
\usepackage{dcolumn}
\usepackage{comment}
\usepackage{hyperref}
\usepackage{natbib}
\usepackage{breqn}
\usepackage{xcolor}

\usepackage[english]{babel}
\usepackage{natbib}

\newcommand{\etal}{\textit{et al}.}

\graphicspath{{./graphs/}}

\makeatletter
\let\cat@comma@active\@empty
\makeatother

\begin{document}

\title{Inflation with antisymmetric tensor field: new candidates}

\author{Sandeep Aashish}
\email{sandeepaashish@klu.ac.in}
\affiliation{Department of Physics, School of Advanced Sciences, Kalasalingam Academy of Research and Education, Krishnankoil, Virudhunagar - 626126, India}

\author{Abhijith Ajith}
\email{abhijith18@iiserb.ac.in}

\author{Sukanta Panda}
\email{sukanta@iiserb.ac.in}

\author{Rahul Thakur}
\email{rahul19@iiserb.ac.in}
\affiliation{Department of Physics, Indian Institute of Science Education and Research, Bhopal - 462066, India}

\date{\today}

\begin{abstract}
    We study classes of inflation models driven by antisymmetric tensor field, with minimal and nonminimal couplings to gravity, that address the known issues of such models considered in the past. First, we show that with a different choice of the background structure of the antisymmetric tensor field, inflation is supported even for the minimal model with quadratic potential contrary to past results. We also include the nonminimal coupling to gravity and analyse perturbations to the antisymmetric tensor as well as the tensor modes of perturbed metric. The two models differ in terms of the behaviour of tensor modes, where the speed of the gravitational wave can be tuned to $c$ in the latter model. The power spectrum and spectral index receive slight scale dependence.  Finally, we consider a quartic potential motivated by the graceful exit to reheating phase, which requires a nonminimal coupling to support inflation. The two tensor modes of the perturbed metric are found to evolve differently in this model, and give rise to a highly scale-dependent power spectrum.
\end{abstract}

\maketitle

\newpage

\section{Introduction} \label{sec:1}
Inflation as a paradigm has been successful in supporting the big bang cosmology, solving problems like the horizon problem, flatness problem and the monopole problem. Further, the structure formation in the universe can be explained by invoking the idea of quantum fluctuations generated in the inflationary era \textcolor{blue}{\cite{Riotto:2002yw}}. Inflationary model building, therefore, has been a subject of strong research in the past decades. Inflation models either employ an external driving field or a modification to gravity \textcolor{blue}{\cite{Inagaki:2019hmm,Zhang:2021ppy,Sangtawee:2021mhz,Bamba:2015uma,Bhattacharjee:2021kar}}. Models employing a single scalar field (generally called the \textit{inflaton}) or multiple scalar fields have been extensively studied in the scientific literature \textcolor{blue}{\cite{Riotto:2002yw,Abedi:2016sks,Kodama:2021yrm,Wands:2007bd,Gong:2006zp,Ohashi:2011na,Vazquez:2018qdg,Bartolo:2021wpt}}. 

The observational data from the CMB is used for testing the authenticity of such models. The recent CMB data by Planck \textcolor{blue}{\cite{Iacconi:2019vgc,Planck:2018jri,Planck:2019kim}} either rules out or applies tight constraints to most conventional models. Moreover, the swampland conjectures in string theory have put additional restrictions on these scalar field models \textcolor{blue}{\cite{Brennan:2017rbf, Andriot:2018mav,Obied:2018sgi,Garg:2018reu,kinney2019,Kallosh:2019axr}}. Despite the abundance of single- and multi-scalar field models of inflation, however, the statistical anomalies in the CMB like anisotropy and dipolar asymmetry remain unexplained \textcolor{blue}{\cite{Planck:2018jri}}.  

This has motivated the exploration of alternative inflation models where the driving field is not a scalar. Several vector and tensor field models have been explored in the past \textcolor{blue} {\cite{Golovnev:2008cf,Darabi:2014aaa,Bertolami:2015wir,Emami:2016ldl,Koh:2009ne,Aashish:2018lhv,Aashish:2019zsy,Aashish:2020mlw,Paul:2020duu,Golovnev:2011yc,Rodriguez:2015rua}}. Vector field models have been shown to suffer from generic ghost and gradient instabilities \textcolor{blue}{\cite{Himmetoglu:2009qi,BeltranJimenez:2017cbn,Golovnev:2011yc}}. Recently however, the study of inflationary dynamics with multiple vector fields has indicated improvements over past models \textcolor{blue}{\cite{Gorji:2020vnh,Murata:2021vnb,jiro2009,jiro2010a}}. In the context of tensor field, several studies of inflationary scenario in the presence of nonsymmetric tensors, in particular the antisymmetric tensor, exist in the literature \textcolor{blue}{\cite{Prokopec:2005fb,Koivisto:2009sd,Obata:2018ilf,Elizalde:2018rmz}}. The appearance of antisymmetric tensor in the early universe is inspired by the superstring models, which give rise to antisymmetric tensor field in the low energy limit \textcolor{blue}{\cite{Rohm:1986,Ghezelbash:2009gf,jiro2010b}}. The inflationary cosmology with $n-$forms was first studied in Refs. \textcolor{blue}{\cite{jiro2010b,jiro2013,jiro2015}}. Rank-2 and rank-3 antisymmetric tensor fields as a standalone inflation-driving field was by Koivisto \textcolor{blue}{\cite{Koivisto:2009ew,Koivisto:2009sd}}, where generic instabilities similar to vector models were highlighted. Later, it was shown by Aashish \etal that inflation is indeed supported by a rank-2 antisymmetric tensor field with non-minimal couplings to gravity \textcolor{blue}{\cite{Aashish:2018lhv}}. This new class of models is free from ghost and gradient instabilities \textcolor{blue}{\cite{Aashish:2019zsy}} and can predict a nearly scale invariant spectrum for tensor perturbations in the quasi de-Sitter limit \textcolor{blue}{\cite{Aashish:2020mlw}}. 

The structure of a rank-2 antisymmetric tensor field, $B_{\mu\nu}$, can be decomposed into two spin-$1$ fields following the SVT decomposition, \textcolor{blue}{\cite{Dodelson:2003ft,Guzzetti:2016mkm}}, similar to the electric and magnetic components of the electromagnetic field strength tensor \textcolor{blue}{\cite{Altschul:2009ae}}. The analysis of Refs. \textcolor{blue}{\cite{Aashish:2018lhv,Aashish:2019zsy,Aashish:2020mlw}} considered a choice of the background structure of $B_{\mu\nu}$ akin to turning off the electric field components of electromagnetic field strength tensor, out of symmetry constraints and calculational convenience, so that $B_{0i}=0$ and $B_{ij}\neq 0$. In this work, we employ a different choice of the background structure of antisymmetric tensor, wherein $B_{ij}=0$ and $B_{0i}\neq 0$. Subsequently the de-Sitter solutions and perturbations to field and metric (excluding scalar and vector modes, for now) have been analysed for stability as well as observational constraints in the context of gravitational wave speed and spectral index. In light of the requirements of a graceful exit to reheating era, we also study a quartic potential which leads to phenomenologically interesting results in the context of distinguishable evolutions of tensor modes. A complete perturbative analysis including the scalar and vector modes is not part of the current work however, and will be addressed in a future work.

The basic requirement for a viable inflationary model is to have stable de-Sitter solutions, with 60-70 e-folds in the slow roll limit \textcolor{blue}{\cite{Dodelson:2003ft}}. The standard choice for the background metric during inflation is the FLRW metric which has the properties of homogeneity and isotropy. Choosing the $(-,+,+,+)$ signature, our metric reads,
\begin{equation}\label{2.1}
     g_{00}=-1, \hspace{.5cm}  g_{ij}=a(t)^{2}\delta_{ij}
 \end{equation}
Owing to the antisymmetry, there are in general six independent components of $B_{\mu\nu}$. 
In this work, our choice of the background antisymmetric tensor $B_{\mu\nu}$ is as follows,
\begin{equation} \label{2.4}
    B_{\mu\nu}=\left ( \begin{array}{cccc}
    0 & B(t) & B(t) & B(t) \\
    -B(t) & 0 & 0 & 0 \\
    -B(t) & 0 & 0 & 0 \\
    -B(t) & 0 & 0 & 0 \\
    \end{array} \right )
\end{equation}
where for calculational convenience we take $B_{0i}(t,\vec{x})=B(t)$ for some scalar field $B(t)$. 
The general model considered in Ref. \textcolor{blue}{\cite{Aashish:2019zsy}} and inspired by the Lorentz violation action studied in Ref. \textcolor{blue}{\cite{Altschul:2009ae}} is given by,
\begin{dmath} \label{2.3}
     S=\int d^{4}x\sqrt{-g}[\frac{R}{2\kappa}-\frac{1}{12}H_{\lambda\mu\nu}H^{\lambda\mu\nu}+\frac{\tau}{2}(\nabla_{\lambda}B^{\lambda\nu})(\nabla_{\mu}{B^{\mu}}_{\nu})-B_{\mu\nu}B^{\mu\nu}(\frac{m^2}{4}-\frac{\xi}{2\kappa}R)+\frac{\zeta}{2\kappa} B^{\lambda\nu}{B^{\mu}}_{\nu}R_{\lambda\mu}]
\end{dmath}
A brief overview of the results of Refs. \textcolor{blue}{\cite{Aashish:2018lhv,Aashish:2019zsy,Aashish:2020mlw}} is in order. The action term with the parameter $\tau$ eliminates the ghost instabilities that occur while perturbing the tensor field $B_{\mu\nu}$ \textcolor{blue}{\cite{Aashish:2019zsy}} in an unperturbed FLRW background metric, atleast for the trivial choice of background structure of $B_{\mu\nu}$. The tensor modes of metric perturbations, which constitute the primordial gravitational waves in this model, give rise to GW speed that can be tuned to the speed of light by controlling the strength of both nonminimal couplings in Eq. (\ref{2.3}) \textcolor{blue}{\cite{Aashish:2020mlw}}. The power spectrum and spectral index are found to be nearly scale invariant. We note that in general, the consistency of the model remains open since all six degrees of freedom are propagating, and continues to be the subject of future investigations. 

In subsequent sections, we will revisit these calculations in the context of the new choice of the background $B_{\mu\nu}$. The organization of this paper is as follows. In section \ref{sec:3} we construct the minimal model and does its perturbative analysis in section \ref{sec:4}. Section \ref{sec:5} deals with the non-minimal model, and the perturbations in this case are analyzed in section \ref{sec:6}. In section \ref{sec:7}, we go through the features of a new potential which is quartic in the driving tensor field. We conclude our findings and address the future possibilities in section \ref{sec:8}.

\section{The Minimal model} \label{sec:3}
Given the choice of background, Eq. (\ref{2.4}), we start our analysis with the minimal model,
\begin{equation} \label{2.5}
     S=\int d^{4}x\sqrt{-g}\left[\frac{R}{2\kappa}-\frac{1}{12}H_{\lambda\mu\nu}H^{\lambda\mu\nu}+\frac{\tau}{2}(\nabla_{\lambda}B^{\lambda\nu})(\nabla_{\mu}{B^{\mu}}_{\nu})-V(B)\right]
\end{equation}
Here $g$ is the metric determinant, $R$ the Ricci tensor for our given spacetime. $\kappa$ is the inverse of squared Planck mass. $\tau$ is a dimensionless parameter. $H_{\lambda\mu\nu}$ is a three ranked tensor defined as, $H_{\lambda\mu\nu}=\nabla_{\lambda}B_{\mu\nu}+\nabla_{\nu}B_{\lambda\mu}+\nabla_{\mu}B_{\nu\lambda}$. This along with the term proportional to $\tau$ constitute the kinetic terms in $B_{\mu\nu}$ for our action \textcolor{blue}{\cite{Altschul:2009ae}}. $V(B)$ is the potential term. We here consider a potential quadratic in $B_{\mu\nu}$ given by $V(B)=m^2 B_{\mu\nu}B^{\mu\nu}/4$. However, for our choice of the background $B_{\mu\nu}$ structure, the term $B_{\mu\nu}B^{\mu\nu}$ is negative. Thus the quadratic potential is always negative. The presence of such a completely negative potential will make our system bounded always, which is undesirable. To get around this we introduce a positive shift $\phi_{0}^{2}$ in our potential. Thus, we have
\begin{equation} \label{2.6}
    V=\frac{m^2}{4}(B_{\mu\nu}B^{\mu\nu}+\phi_{0}^{2})
\end{equation}
An inflationary theory should be able to support stable de-Sitter solutions. In order to ensure this, we begin by determining the Einstein solutions for our minimal model. They are found by varying the action with respect to the metric tensor $g_{\mu\nu}$. In tensor form, the Einstein equation reads,
\begin{equation} \label{3.1}
    G_{\mu\nu}=\kappa T_{\mu\nu}
\end{equation}
where $G_{\mu\nu}$ is the Einstein tensor which arises solely from the Einstein-Hilbert part of the action. It has the following form in our setup.
\begin{equation} \label{3.2}
    G_{00}=3H^2 \hspace{1cm} G_{0i}=0 \hspace{1cm} G_{ij}=-a^2(2\dot{H}+3H^2)
\end{equation}
Here, $H$ is the Hubble parameter defined as $H\equiv H(t)=\dot{a}(t)/a(t)$. We can rewrite the stress-energy tensor $T_{\mu\nu}$ in terms of the contributions coming individually from the $\tau$ term, and the rest of the terms containing $B_{\mu\nu}$ for convenience, i.e $T_{\mu\nu}=T_{\mu\nu}^{\tau}+T_{\mu\nu}^M$.  We adopt their tensorial expressions from Refs. \textcolor{blue}{\cite{Aashish:2018lhv}} and \textcolor{blue}{\cite{Aashish:2019zsy}}.

\begin{equation} \label{3.3}
    T_{\mu\nu}^M=\frac{1}{2}{H^{\alpha\beta}}_{\mu}H_{\nu\alpha\beta}+m^2{B^{\alpha}}_{\mu}B_{\alpha\nu}-g_{\mu\nu}(\frac{1}{12}H_{\alpha\beta\gamma}H^{\alpha\beta\gamma} +\frac{1}{4}m^2B_{\alpha\beta}B^{\alpha\beta}-\frac{m^2}{4}\phi_{0}^2)
\end{equation}
\begin{dmath} \label{3.4}
    T_{\mu\nu}^{\tau}=\frac{\tau}{2}[g_{\mu\nu}((\nabla_{\lambda}B^{\sigma\lambda})(\nabla_{\rho}{B^{\rho}}_{\sigma})+2B^{\sigma\lambda}\nabla_{\lambda}\nabla_{\rho}{B^{\rho}}_{\sigma})+2(\nabla_{\lambda}{B^{\lambda}}_{\mu})(\nabla_{\rho}{B^{\rho}}_{\nu})\\+2({B_{\mu}}^{\lambda}\nabla_{\lambda}\nabla_{\rho}{B_{\nu}}^{\rho}+{B_{\nu}}^{\lambda}\nabla_{\lambda}\nabla_{\rho}{B_{\mu}}^{\rho})]
\end{dmath}
The tensor equation, Eq. (\ref{3.1}), after exploiting the functional forms of the expressions and simplifying further, we get a system of two equations. We now define,
\begin{equation} \label{3.5}
    B(t)=a(t)\phi(t)
\end{equation}
We use this redefinition to simplify our equations and to obtain a form of equations similar to that of a scalar field model. We have,
\begin{equation} \label{3.6}
    H^{2}=\kappa m^{2}\phi^{2}-\kappa\tau(H\phi\dot{\phi}+2H^{2}\phi^{2})+\frac{\kappa m^{2}}{12}\phi_{0}^{2}
\end{equation}

\begin{equation} \label{3.7}
    2\dot{H}+3H^{2}=3\kappa\tau(H\phi\dot{\phi}+2H^{2}\phi^{2})+\frac{\kappa m^{2}}{4}\phi_{0}^{2}
\end{equation}

\subsection{The de-Sitter solutions }
In any inflationary model, at the onset of inflation, the spacetime can be described by the FLRW metric with certain preconditions. One of the preconditions is the presence of de-Sitter solutions. The spacetime will start to evolve from these solutions. By definition, the Hubble parameter $H$ can be treated as a constant during the exponential expansion of the universe. Furthermore, during inflation the field $\phi$ evolves from the de-Sitter value and goes through the slow rolling phase for about 60-70 e-folds. During slow roll, the value of the field does not vary considerably. So, the field $\phi$ can be roughly treated as a constant while determining the de-Sitter solutions. We look whether our model can permit such solutions. Applying the constraints $\dot{\phi}\sim\dot{H}\sim 0$ to our system of equations, Eq. (\ref{3.6}) and Eq. (\ref{3.7}), we get the following solutions.
\begin{equation} \label{3.9}
    H_{d}^{2}=\frac{m^{2}}{4\tau}, \hspace{2em}  \phi_{d}^{2}=\frac{1}{2\kappa\tau}-\frac{\phi_{0}^{2}}{6}
\end{equation}
The positivity of these solutions can be ensured by demanding that,
\begin{equation} \label{3.10}
    \tau>0, \hspace{2em}  \kappa\tau\phi_0^2<3
\end{equation}
It is interesting to note that for our choice of $B_{\mu\nu}$, given in Eq. (\ref{2.4}), de-Sitter solutions can be obtained without the need of any non-minimal couplings with gravity unlike past studies \cite{Aashish:2018lhv} where de-Sitter solutions required the presence of non-minimal couplings. Thus, the new choice can give such solutions in a more simple scenario. The stability of the de-Sitter solutions must be checked. For that, we can perturb our system around the de-Sitter values. If the perturbations are to decay in time, we can say that the de-Sitter solutions are stable. So we have,
\begin{equation} \label{3.11}
    H=H_d+\delta H \hspace{0.5cm} \phi=\phi_d+\delta\phi
\end{equation}
Here, $\delta H$ and $\delta\phi$ are the linear order perturbations around the de-Sitter background. We substitute them in Eqs. (\ref{3.6}) and (\ref{3.7}). At linear order, we have,
\begin{equation} \label{3.12}
    \dot{\delta\phi}=\left(\frac{2m^2}{\tau H_d}-4H_d\right)\delta\phi-\left(\frac{2}{\kappa\tau\phi_0}+4\phi_d\right)\delta H
\end{equation}
\begin{equation} \label{3.13}
    \dot{\delta H}=3\kappa m^2\phi_d\delta\phi-6H_d\delta H
\end{equation}
Eqs. (\ref{3.12}) and (\ref{3.13}) can be recast into a matrix equation of the form,
\begin{equation}\label{3.14}
    \dot{\Xi}=\Lambda\Xi
\end{equation}
where,
\begin{equation} \label{3.15}
    \Xi=\begin{pmatrix}
    \delta\phi \\
    \delta H \\
    \end{pmatrix}, \hspace{1cm} \Lambda=\begin{pmatrix}
    \frac{2m^2}{\tau H_d} -4H_{d}&-(4\phi_{d} +\frac{2}{\kappa\tau\phi_{d}}) \\
     3\kappa m^2\phi_d & -6H_{d}  \\
     \end{pmatrix}
\end{equation}
The solution of Eq. (\ref{3.14}) has the general form,
\begin{equation} \label{3.16}
    \Xi=Ce^{\lambda_1t}+De^{\lambda_2t}
\end{equation}
where $C$ and $D$ are coefficient matrices determined by the boundary conditions. $\lambda_1$ and $\lambda_2$ are the eigenvalues of the matrix $\Lambda$.
The eigenvalues can be calculated from the trace and determinant of $\Lambda$.
\begin{equation} \label{3.17}
    \lambda_1 +\lambda_2=\left(\frac{2m^2}{\tau H_{d}^2} -10\right)H_{d}=-2H_d
\end{equation}
\begin{equation} \label{3.18}
    \lambda_1 \lambda_2=H_d^2\left(24-\frac{6m^2}{\tau H_d^2}+\frac{12\kappa m^2\phi_d^2}{H_d^2}\right)=12\kappa m^2\phi_d^2
\end{equation}
Solving we get,
\begin{equation}\label{3.19}
    \lambda_{1(2)}=-H_dt+(-)i\sqrt{23-8\kappa\tau\phi_0^2}H_dt
\end{equation}
So,
\begin{equation} \label{3.20}
    \Xi=e^{-H_dt}(Ce^{i\sqrt{23-8\kappa\tau\phi_0^2}H_dt}+De^{-i\sqrt{23-8\kappa\tau\phi_0^2}H_dt})
\end{equation}
In the following section, it will be evident that slow roll conditions are valid in the limit where $\kappa\tau\phi_0^2$ takes very small positive values. In this limit, in Eq. (\ref{3.20}), the quantity inside the square root is positive. So the perturbations are oscillating with a constant damping in time. Thus the perturbations are decaying and the de-Sitter background turns out to be stable.

\subsection{The slow roll parameters}
The de-Sitter solutions correspond to an ideal scenario where $H$ and $\phi$ are strictly constant. Now, we consider a nearly de-Sitter spacetime, where we relax the strictness on the constancy of $H$ and $\phi$. For a successful inflationary scenario, the model should be able to support at least 70 e-folds of inflation \textcolor{blue}{\cite{Dodelson:2003ft}}. We try to implement this condition through the slow roll parameters. The first slow roll parameter we choose is $\epsilon$ which controls the acceleration of the universe.
\begin{equation}\label{3.21}
     \epsilon=-\frac{\dot{H}}{H^2} \hspace{1em} \Longrightarrow \hspace{1em} \frac{\ddot{a}}{a}=H^2(1-\epsilon)
\end{equation}
So for a positive acceleration, the slow roll parameter has to be small. An $\epsilon$ value of 1 marks the end of inflation. We choose $\delta$ as the second slow roll parameter \textcolor{blue}{\cite{Aashish:2018lhv}}, defined as,
\begin{equation} \label{3.22}
    \delta=\frac{\dot{\phi}}{\phi H}
\end{equation}
This parameter is directly related to the flatness of our quadratic potential $V(\phi)$ which drives inflation. The definitions of the slow roll parameters $\epsilon$ and $\delta$ contain first order time derivatives. The slow roll parameters in the generic scalar field inflation models also constitute second order time derivatives. Our theory being a tensor field theory has a complicated background structure. So, we are looking for a simple scenario during slow roll analysis wherein, we consider only the slow roll parameters with first order time derivatives. Higher order slow roll parameters will be discussed separately in a future work. From Eqs. (\ref{3.6}) and (\ref{3.7}), we have,
\begin{equation} \label{3.23}
    2\dot{H}=6\kappa\tau(H\phi\dot{\phi}+2H^2\phi^2)-3\kappa m^2\phi^2
\end{equation}
Using the definition of slow roll parameters given in Eqs. (\ref{3.21}) and (\ref{3.22}), we can write Eq. (\ref{3.23}) as,
\begin{equation} \label{3.24}
    -2\epsilon=6\kappa\tau\phi^2(\delta+2) -3\kappa m^2\frac{\phi^2}{H^2}
\end{equation}
Assuming the values of $\phi^2$ and $H^2$ doesn't vary considerably from their de-Sitter counterparts, we have,
\begin{equation} \label{3.25}
    \delta\approx -\frac{2\epsilon}{3(1-\frac{\theta}{3})}
\end{equation}
where $\theta=\kappa\tau\phi_0^2$. The values of $\tau$ and $\phi_0^2$ should be positive according to the ghost free conditions and the positivity requirement of de-Sitter solutions. The parameters $\tau$ and $\phi_0^2$ can be tuned such that $\theta<<1$. Then, we can see from Eq. (\ref{3.25}) that $\epsilon$ and $\delta$ are of the same order. The smallness of $\delta$ will ensure that the driving potential $V(\phi)$ is sufficiently flat. In addition, this will ensure that $\epsilon$ is small, satisfying all the criteria for slow roll inflation. Also, we can relate the number of e-folds to the slow roll parameter $\delta$ by the following relation, 
\begin{equation} \label{3.26}
    N=\int^{t_f}_{t_i}Hdt=\int^{\phi_f}_{\phi_i}\frac{H}{\dot{\phi}}d\phi=\frac{1}{\delta}\int^{\phi_f}_{\phi_i}\frac{1}{\phi}d\phi=\frac{1}{\delta}ln\left(\frac{\phi_f}{\phi_i}\right)
\end{equation}
Here, $\delta$ is the only parameter controlling the number of e-folds, which appears in the denominator. Hence, the smallness of $\delta$ will ensure enough number of e-folds during inflation. We can write, 
\begin{equation} \label{3.27}
    \delta<\frac{1}{70}ln\left(\frac{\phi_f}{\phi_i}\right)
\end{equation}
We have to note that after giving the substitution in Eq. (\ref{3.5}) our quadratic potential reads as,
\begin{equation} \label{3.8}
    V(\phi)=\frac{m^2}{4}(\phi_0^2-6\phi^2)
\end{equation}
This potential can support the slow rolling down a potential hill and is an example of the Hill-top inflation models discussed in the scientific literature \textcolor{blue}{\cite{Boubekeur:2005zm,Enckell:2018kkc,Kohri:2007gq}}. Though, inflation can be supported by this potential, a problem occurs for the reheating scenario occurring after the inflationary period. Reheating is the phase in which the energy carried by the driving field gets converted to standard model particles. In the scalar field theory, the driving inflaton field oscillates around the effective potential minimum during this phase \textcolor{blue}{\cite{Kofman:1994rk}}. In analogy, the form of our potential does not provide a minimum and a graceful exit to the reheating phase is not naturally possible. A potential resolution to this problem is to work with a different choice of potential, which is the subject of Sec. \ref{sec:7}.

\section{Minimal model: Perturbations} \label{sec:4}
In the previous sections, we were able to see that our model can support stable de-Sitter solutions and can admit slow roll conditions. Now, we should develop the cosmological perturbation theory for our model. The perturbations can simultaneously arise from both the metric side as well as from the inflation driving field $B_{\mu\nu}$. Following the SVT decomposition, \textcolor{blue}{\cite{Dodelson:2003ft,Guzzetti:2016mkm}}, the metric perturbations can be expressed in the following form,
\begin{equation} \label{4.1}
\begin{array}{c}
    \delta g_{00}=-\psi \hspace{1cm} \delta g_{0i}=a(\partial_i \chi +E_i) \\
    \vspace{0.1mm}\\
    \delta g_{ij}=a^2(-2\alpha\delta_{ij}+2\partial_{ij}\beta+(\partial_iF_j+\partial_jF_i)+h_{ij})
    \end{array}
\end{equation}
where $\psi$,$\chi$,$\alpha$ and $\beta$ are scalar modes, $E_i$ and $F_i$ divergence free vector modes, and $h_{ij}$ constitutes the traceless, transverse tensor modes. By the decomposition theorem, the metric scalar, tensor and vector modes separate at linear order and can be studied individually. Initially, we aim to study only the tensor perturbations coming from the metric. The perturbations from $B_{\mu\nu}$ take the following form,
\begin{equation} \label{4.2}
    \delta B_{0i}=-E_i \hspace{2em} \delta B_{ij}=\epsilon_{ijk} M_k
\end{equation}
where $\epsilon_{ijk}$ is the completely antisymmetric Levicivita tensor of rank 3. These perturbations can be decomposed in the following manner.
\begin{equation} \label{4.3}
    \Vec{E}=\Vec{\nabla}u+\Vec{U}, \hspace{1cm} \Vec{M}=\Vec{\nabla}v+\Vec{V}
\end{equation}
where $\Vec{U}$ and $\Vec{V}$ are divergence free vector fields. We can see that this structure doesn't permit tensor modes. Hence, the $B_{\mu\nu}$ perturbations and the tensor perturbations in metric can be studied separately. First, we study the $B_{\mu\nu}$ perturbations. We follow the perturbative analysis for the $B_{\mu\nu}$ modes performed in Ref. \textcolor{blue}{\cite{Aashish:2019zsy}} to check for ghost instabilities. Ghost instabilities occur when the coefficients of the kinetic terms of the perturbed modes acquire negative coefficients. They make the theories ill defined and cause the energy to be unbounded from below. The perturbed action is analyzed in Fourier space to eliminate the spatial derivatives. The coefficient matrix for the kinetic terms will look like,
\begin{equation} \label{4.4}
    T=diag\begin{bmatrix}
    \frac{k^2}{2a(t)} & \frac{1}{2a(t)} & \frac{1}{2a(t)} & \frac{k^2a(t)\tau}{2} & \frac{a(t)\tau}{2} & \frac{a(t)\tau}{2} \\
    
    \end{bmatrix}
\end{equation}
The matrix has positive eigen values, given the condition $\tau>0$. But this constraint is already accomplished in our model. The kinetic terms have positive coefficients and from this analysis, the theory is free from ghost instabilities. We again note that the ghost-free conditions here are subject to the conditionalities akin to the results of Ref. \textcolor{blue}{\cite{Aashish:2019zsy}}, namely our choice of background $B_{\mu\nu}$, the metric, and considering only the perturbations in the background field $B_{\mu\nu}$. We can expect pathological instabilities in a more general setup since all the six degrees of freedom are propagating in general, and will be the subject of future studies. Also, the complete perturbative analysis would involve the scalar and vector perturbations to the metric as well and will be addressed in upcoming works as the analysis will be more involved. Now, we look at the tensor modes from the metric. The metric tensor reads,
\begin{equation} \label{4.5}
    g_{00}=-1, \hspace{1cm} g_{0i}=0, \hspace{1cm} g_{ij}=a^2(\delta_{ij}+h_{ij})
\end{equation}
The tensor perturbation $h_{ij}$ is gauge invariant when we consider the perturbation theory at linear order \textcolor{blue}{\cite{Malik:2008im}}. These tensor modes can manifest in the form of gravitational waves. Such primordial gravitational waves have been predicted in many inflationary models \textcolor{blue}{\cite{Abbott:1984fp,Rubakov:1982df,Fabbri:1983us}}. Thus they are believed to contain the signatures regarding the early universe physics and are of great interest. The following form for $h_{ij}$ is employed satisfying its conditions of traceless and non-transverse nature, \textcolor{blue}{\cite{Dodelson:2003ft}},
\begin{equation} \label{4.6}
    h_{ij}=\begin{pmatrix}
    h_+ & h_{\times} & 0 \\
    h_{\times} & -h_+ & 0\\
    0 & 0 & 0 
    \end{pmatrix}
\end{equation}
Here $h_+$ and $h_{\times}$ lie in the X-Y plane, whereas the wave vector $\Vec{k}$ is oriented in $z$ direction.
We apply these values to our action in Eq. (\ref{2.5}). We analyse the action at second order in these perturbations. This is because the kinetic terms arise at second order in perturbations. Also, the second order action gives rise to gravitational waves \textcolor{blue}{\cite{Guzzetti:2016mkm}}. The second order part reads as,
\begin{dmath} \label{4.7}
S_2=\int dtd^3x \frac{a^3}{4\kappa}[(\dot{h_{\times}}^2+\dot{h_{+}}^2)(1+4\kappa\tau\phi^2)+(h_{\times}\dot{h_{\times}}+h_+\dot{h_+})(4\kappa\tau(\phi\dot{\phi}+2H\phi^2)-4H)+(h_{\times}^2+h_+^2)(\kappa[m^2(\phi^2+\frac{\phi_0^2}{2})+\tau(\dot{\phi}+2H\phi)^2]-6(\dot{H}+2H^2))+3(h_{\times,z}^2+h_{+,z}^2)+4(h_{\times}h_{\times,zz}+h_{+}h_{+,zz})]+S_{xy}
\end{dmath}
where $S_{xy}$ consists of terms containing partial derivatives of the tensor modes with respect to the position coordinates-$x$ and $y$. At this stage, it is convenient to work in Fourier space. We choose the momentum vector $\Vec{k}$ to lie in the $Z$ direction.
\begin{equation} \label{4.8}
    f(\Vec{x},t)=\int d^3k e^{-ikz}\Tilde{f}(\Vec{k},t)
\end{equation}
From now, the notations for the tensor modes will actually be representing their Fourier transforms. We omit the tilde in the actual symbols for simplicity. Now, $S_{xy}$ will go to zero. We can see that the tensor modes $h_+$ and $h_{\times}$ decouple, and the action can be written as a summation. Thus the Fourier transformed action will look like,
\begin{equation} \label{4.9}
    S_2^{FT}=\sum_{e=+,\times}\int dtd^3k\frac{a^3}{4\kappa}[\Omega_k\dot{h_e}^{\dagger}\dot{h_e}+\Omega_c(\dot{h_e}^{\dagger}h_e+h_e^{\dagger}\dot{h_e})+\Omega_gh_e^{\dagger}h_e]
\end{equation}
where the coefficients are given by,
\begin{equation}\label{4.10}
\begin{array}{c}
    \Omega_k=1+4\kappa\tau\phi^2, \hspace{1cm} \Omega_c=-2H+2\kappa\tau(2H\phi^2+\phi\dot{\phi}) \\
    \vspace{0.1mm}\\
    \Omega_g=\kappa m^2\left(\phi^2+\frac{\phi_0^2}{2} \right)-6(\dot{H}+2H^2)+\kappa\tau(\dot{\phi}+2H\phi)^2-\frac{k^2}{a^2}
    \end{array}
\end{equation}

\begin{figure}[h!]
    \centering
    \includegraphics[width=.7\linewidth,height=.4\linewidth]{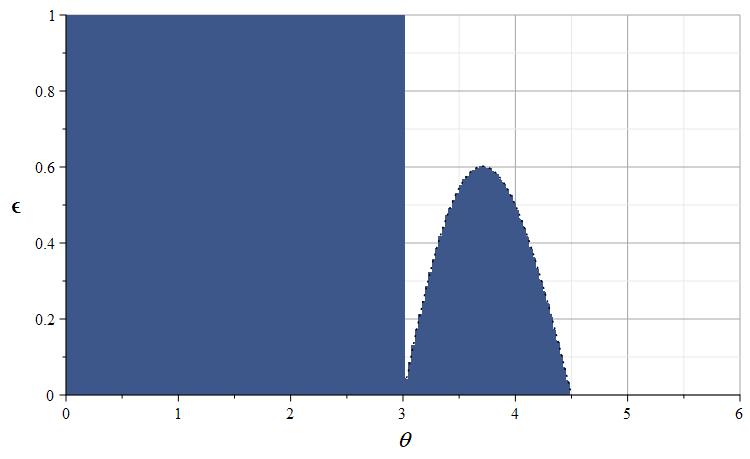}
    \caption{The shaded region shows the allowed values of $\theta$ and $\epsilon$ satisfying the condition $\Omega_k>0$}
    \label{fig:1}
\end{figure}

To eliminate ghost instabilities, the coefficient $\Omega_k$ must be positive. In the quasi de-Sitter limit, we can approximate,
\begin{equation} \label{4.11}
    \kappa\tau\phi^2\approx\frac{3-\theta}{6}+\epsilon\left(\frac{6-\theta}{18} \right)+\delta\left(\frac{3-2\theta}{36} \right)+O(\epsilon^2,\delta^2,\epsilon\delta)
\end{equation}
So, we have,
\begin{equation} \label{4.12}
    \Omega_k\approx3(1-\frac{2\theta}{9})+\epsilon\left(\frac{12-2\theta}{9} \right)+\delta\left(\frac{3-2\theta}{9} \right)
\end{equation}
$\delta$ can be approximated in terms of $\epsilon$ using Eq. (\ref{3.25}). $\epsilon$ can take values between 0 and 1. The allowed values of $\theta$ satisfying the positivity of $\Omega_k$ is shown in Fig. (\ref{fig:1}).
We already have $\theta<<1$. In this limit, we can see that ghost instabilities are absent. These tensor perturbations propagate in the form of gravitational waves. Instabilities arise when their speed of propagation becomes negative. So, we perform an estimate of the velocity. The equation of motion of these tensor modes can be obtained by varying the perturbed action, Eq. (\ref{4.9}), with respect to $h_e^{\dagger}$. This reads as,
\begin{equation} \label{4.13}
    \ddot{h_e}+\left(\frac{\dot{\Omega_k}}{\Omega_k}+3H\right)\dot{h_e}+\left(\frac{\dot{\Omega_c}+3H\Omega_c-\Omega_g}{\Omega_k}\right)h_e=0
\end{equation}
In the quasi de-Sitter limit, we can write,
\begin{equation} \label{4.14}
    \frac{\dot{\Omega_k}}{\Omega_k}=\frac{8\kappa\tau\phi^2}{1+4\kappa\tau\phi^2}H\delta, \hspace{1em} \frac{\dot{\Omega_c}+3H\Omega_c-\Omega_g}{\Omega_k}=Q_mH^2+\frac{k^2}{a^2}P_m
\end{equation}
where,
\begin{equation} \label{e1}
    Q_m= \frac{8\kappa\tau\phi^2(\delta+2)}{1+4\kappa\tau\phi^2}, \hspace{2em} P_m=\frac{1}{1+4\kappa\tau\phi^2}
\end{equation}
We substitute a wave solution of the form $h_e=A\exp[-i\int^t(c_Tk/a(t'))dt']\Vec{e}$ in Eq. (\ref{4.13}). Here $c_T$ is the velocity of propagation of the gravitational wave, $A$ is a constant and $\Vec{e}$ is a constant vector. We get the dispersion relation as,
\begin{equation} \label{4.15}
    c_T^2+i\left(\frac{2aH}{k}\right)\left(1+\frac{6-2\theta}{9-2\theta} \right)c_T-\left(\frac{aH}{k}\right)^2Q_m-P_m=0
\end{equation}
In the deep subhorizon limit, when $k>>aH$, we can neglect the terms with $\frac{aH}{k}$. So,
\begin{equation} \label{4.16}
    c_T^2=P_m\approx\frac{1}{3(1-\frac{2}{9}\theta)}+\epsilon\left(\frac{-12+2\theta}{(9-2\theta)^2} \right)+\delta\left(\frac{-3+2\theta}{(9-2\theta)^2} \right)
\end{equation}
For our analysis, we have $\theta<<1$. In this limit, the coefficients of the slow roll parameters calculated from Eq. (\ref{4.16}) are of the order of $10^{-1}-10^{-2}$. The smallness of the slow roll parameters further minimize their contributions. Thus, $c_T^2$ is largely dominated by the de-Sitter part which is roughly equal to 1/3. So, we can see that the squared wave velocity acquires a positive value, and is free from instabilities. The model predicts a wave velocity of about $0.577c$, where $c$ is the velocity of light in vacuum. At present, we do not have much experimental evidence regarding the detection of primordial gravitational waves. So the possibility of such a value for the GW velocity cannot be entirely ruled out, though the recent observation from neutron star merger data GW170817 has shown that GWs propagate with the velocity of light in vacuum \textcolor{blue}{\cite{Odintsov:2019clh}}. The GW velocity can be matched with the velocity of light in vacuum if non-minimal coupling terms are incorporated into our initial action. We tend to explore this possibility.

\section{ The $RB_{\mu\nu}B^{\mu\nu}$ coupling} \label{sec:5}
\begin{figure}[h!]
    \centering
    \includegraphics[width=.6\linewidth,height=.4\linewidth]{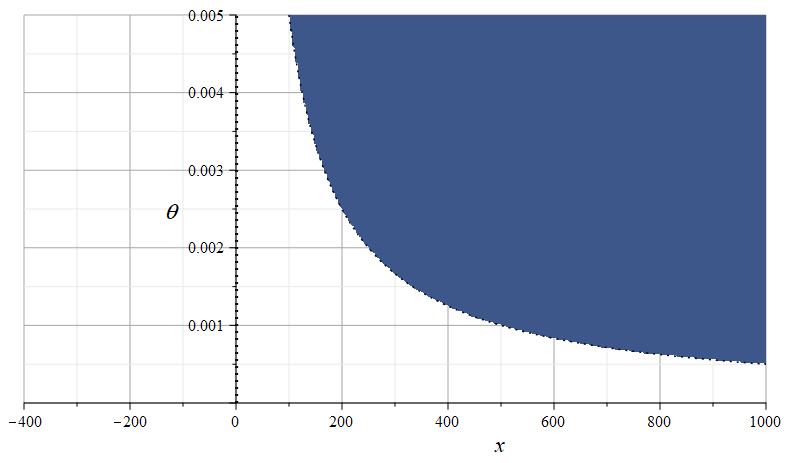}
    \caption{The shaded region indicates the allowed values of $\theta$ and $x$ from the positivity of de-Sitter solutions.}
    \label{fig:3}
\end{figure}

\begin{figure}[h!]
    \centering
    \includegraphics[width=.7\linewidth,,height=.4\linewidth]{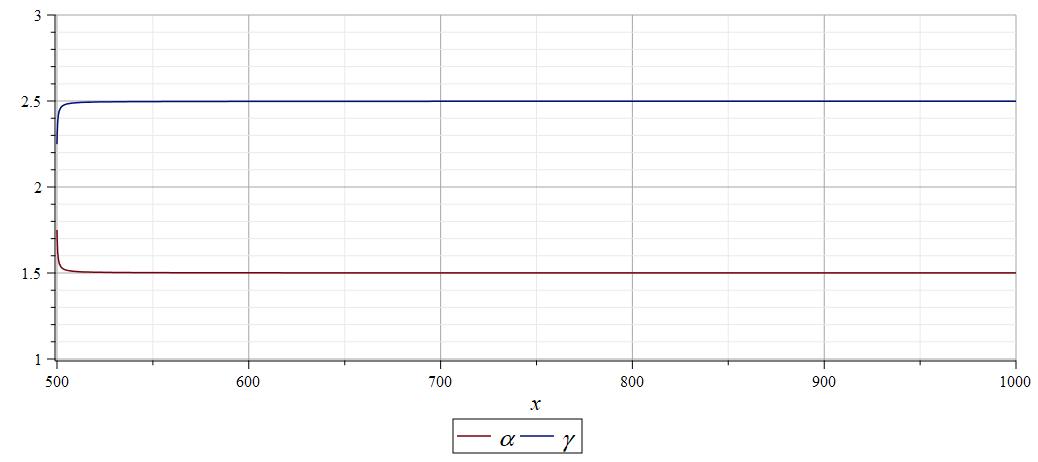}
    \caption{$\alpha$ and $\gamma$ values against $x$ for a $\theta$ value 0.001}
    \label{fig:4}
\end{figure}

We can now add non-minimal couplings with gravity to our initial action in Eq. (\ref{2.5}). In the past models, the addition of non-minimal couplings gave rise to de-Sitter solutions. Further, the primordial GW velocity was tuned to the velocity of light in vacuum using the coupling strengths \textcolor{blue}{\cite{Aashish:2020mlw}}. This motivates us to look at the effects of the non-minimal couplings in our model. The antisymmetric tensor field $B_{\mu\nu}$ can couple with the Riemann curvature tensor, Ricci tensor and Ricci scalar. The general form of these couplings is given in \textcolor{blue}{\cite{Altschul:2009ae}}. Initially, we are adding only the coupling between $B_{\mu\nu}$ and the Ricci scalar $R$, which is simpler than the other two. This is achieved through the term $RB_{\mu\nu}B^{\mu\nu}$ in the action \textcolor{blue}{\cite{Aashish:2018lhv}}. The strength of this coupling is controlled by the parameter $\xi$ which has the dimensions of $M_{Pl}^{-2}$. The action is now given as,
\begin{equation} \label{5.1}
     S=\int d^{4}x\sqrt{-g}\left[\frac{R}{2\kappa}-\frac{1}{12}H_{\lambda\mu\nu}H^{\lambda\mu\nu}+\frac{\tau}{2}(\nabla_{\lambda}B^{\lambda\nu})(\nabla_{\mu}{B^{\mu}}_{\nu})-B_{\mu\nu}B^{\mu\nu}(\frac{m^2}{4}-\frac{\xi}{2\kappa}R)-\frac{m^2\phi_0^2}{4}\right]
\end{equation}
The contribution to Stress-energy tensor due to the new term in the action is given by \textcolor{blue}{\cite{Aashish:2018lhv}},
\begin{dmath} \label{5.2}
    T_{\mu\nu}^{\xi}=\frac{\xi}{\kappa}[\nabla_{\mu}\nabla_{\nu}(B_{\alpha\beta}B^{\alpha\beta})-g_{\mu\nu}\nabla^{\lambda}\nabla_{\lambda}(B_{\alpha\beta}B^{\alpha\beta})-G_{\mu\nu}(B_{\alpha\beta}B^{\alpha\beta})-2R{B^{\alpha}}_{\mu}B_{\alpha\nu}]
\end{dmath}
The Einstein equations are now,
\begin{dmath} \label{5.3}
    H^2=\kappa m^2\phi^2-\kappa\tau(H\phi\dot{\phi}+2H^2\phi^2)+\frac{\kappa m^2\phi_0^2}{12}-6\xi(3\dot{H}\phi^2-2H\phi\dot{\phi}+5H^2\phi^2)
\end{dmath}
\begin{dmath} \label{5.4}
    2\dot{H}+3H^2=3\kappa\tau(H\phi\dot{\phi}+2H^2\phi^2)+\frac{\kappa m^2\phi_0^2}{4}+6\xi(2\phi\ddot{\phi}+2\dot{\phi}^2+4H\phi\dot{\phi}-\dot{H}\phi^2-3H^2\phi^2)
\end{dmath}
The de-Sitter solutions get modified into,
\begin{equation} \label{5.5}
    H_d^2=\frac{m^2}{4\tau(1+6x)}, \hspace{2em}  \phi^2_d=\frac{1}{2\kappa\tau(1-3x)}-\frac{\phi_0^2}{6}\left(\frac{1+6x}{1-3x}\right)
\end{equation}
where $x$ is a dimensionless parameter given by,
\begin{equation} \label{5.6}
   x=\frac{\xi}{\kappa\tau} 
\end{equation}
From here onwards we will be using this parameter instead of $\xi$ in our equations. $H_d^2$ can be made positive by demanding $x$ to be greater than $-1/6$. The positivity of $\phi_d^2$ depends on $\phi_0^2$ as well.
Analysis of the stability of the de-Sitter solutions follows the same recipe as with the minimal model. The fluctuations in $H$ and $\phi$ can be written as,
\begin{equation} \label{5.7}
    \Xi=e^{-\alpha H_dt}(Ce^{\gamma  H_dt}+De^{-\gamma H_dt})
\end{equation}
where $\gamma=\sqrt{\alpha^2-\beta}$. The functions $\alpha$ and $\beta$ are given as,
\begin{dmath} \label{5.8}
    \alpha(x,\theta)=\frac{-1296x^3-1080x^2+96x+4+\theta(2592x^4+3168x^3+48x^2-68x)}{-864x^3-576x^2+24x+4+\theta(1728x^4+1728x^3+72x^2-28x)}
\end{dmath}
\begin{dmath} \label{5.9}
\beta(x,\theta)=\frac{8(72x^3-48x^2+2x+2)(6-\theta(12x+2))}{-864x^3-576x^2+24x+4+\theta(1728x^4+1728x^3+72x^2-28x)}
\end{dmath}
where the parameter $\theta=\kappa\tau\phi_0^2$. We have chosen $\theta$ to have a value of $10^{-3}$. Why we select this value will be clear when we talk about the tensor perturbations in metric. The $x$ values consistent with the positivity of de-Sitter solutions can be identified from Fig. (\ref{fig:3}).
From Fig. (\ref{fig:3}), we choose the value of $x$ to be greater than 500. The behaviour of the functions $\alpha$ and $\gamma$ in this range is shown in Fig. (\ref{fig:4}).
We can see that $\alpha$ and $\gamma$ are taking positive values for our range of $x$. Since, $\alpha$ and $\gamma$ are positive, from Eq. (\ref{5.7}), if we let $C$ go to zero, we will have an exponentially decaying solution for the fluctuations around the de-Sitter point. Hence, we can state the stability of the de-Sitter solutions. Though the mechanism to reach such a solution is not known, we pose this argument for a decaying solution leaving the underlying mechanism for future study.\\
\\
Using the definition of the slow roll parameters $\epsilon$ and $\delta$ defined for the minimal model, we can combine the equations (\ref{5.3}) and (\ref{5.4}) to obtain
\begin{dmath} \label{5.10}
    -2\epsilon=-\frac{3\kappa m^2\phi^2}{H^2}+6\kappa\tau\phi^2(\delta+2)+6x\kappa\tau\phi^2(\frac{2\ddot{\phi}}{H^2\phi}+2\delta^2-2\delta-8\epsilon+12)
\end{dmath}
Ignoring the second order terms, we can write,
\begin{equation} \label{5.11}
    \delta=-\epsilon\left[\frac{2-30x+\theta(8x+48x^2)}{(1-2x)(3-\theta(1+6x))}\right]
\end{equation}
\begin{figure}[h!]
    \centering
    \includegraphics[width=.7\linewidth,height=.4\linewidth]{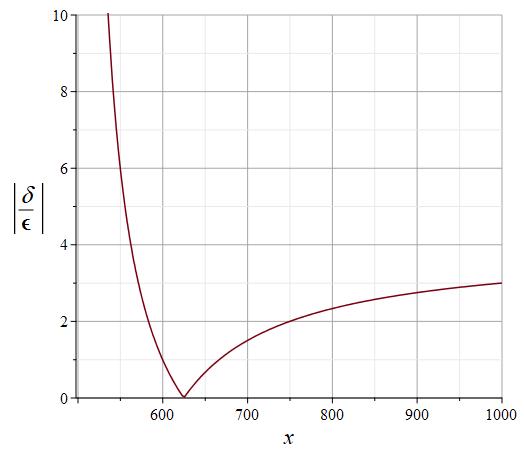}
    \caption{Ratio of slow roll parameter values against $x$ for a $\theta$ value of 0.001}
    \label{fig:5}
\end{figure}
\\
The absolute value of the $\delta$ to $\epsilon$ ratio in our parameter range is given in Fig. (\ref{fig:5}). From the graph, we can see that this ratio decreases initially as we increase the value of $x$ in our parameter range. The function has a zero for $x\approx624.8$. After that, the value rises but gets slower as $x$ becomes larger. From Fig. (\ref{fig:5}), we can safely select $x$ anywhere between 700 and 1000 so that $\epsilon$ and $\delta$ are of the same order. $x$ can take any of these values. Hence, similar to the case of minimal model, we can ensure slow roll conditions with enough number of e-folds. 
\subsection{Perturbations } \label{sec:6}
\begin{figure}[h!]
    \centering
    \begin{subfigure}{.5\textwidth}
    \centering
    \includegraphics[width=.9\linewidth,height=.6\linewidth]{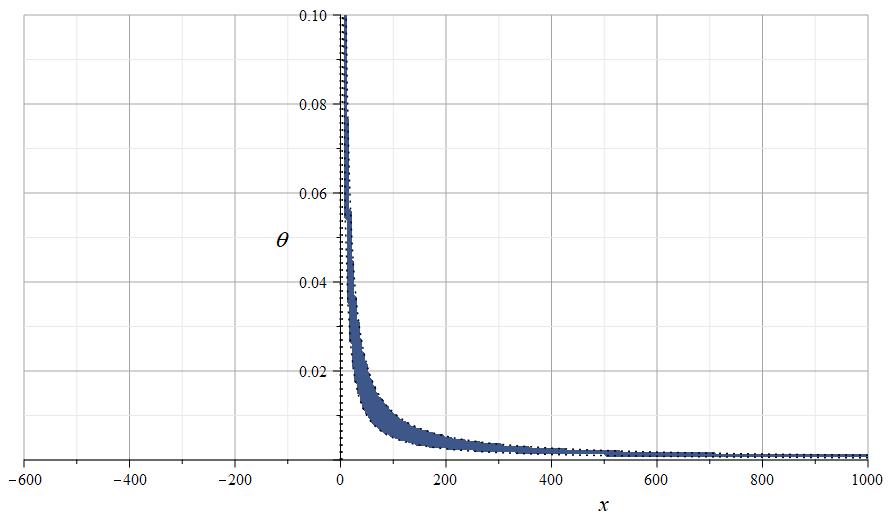}
    \caption{$\epsilon=0.01$}
    \end{subfigure}%
    \begin{subfigure}{.5\textwidth}
    \centering
    \includegraphics[width=.9\linewidth,height=.6\linewidth]{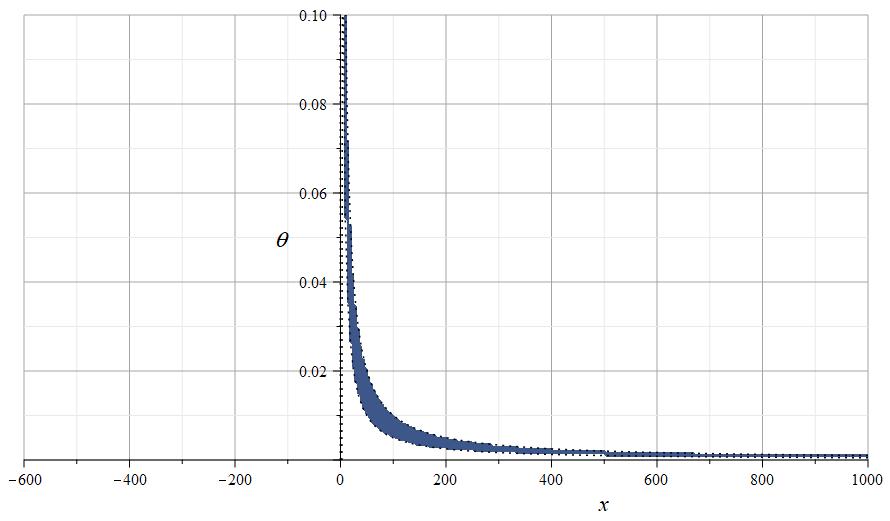}
    \caption{$\epsilon=0.001$}
    \end{subfigure}
    \caption{The shaded region indicates the allowed values of $x$ and $\theta$.}
    \label{fig:7}
\end{figure}

\begin{figure}[h!]
    \centering
    \includegraphics[width=.7\linewidth,height=.4\linewidth]{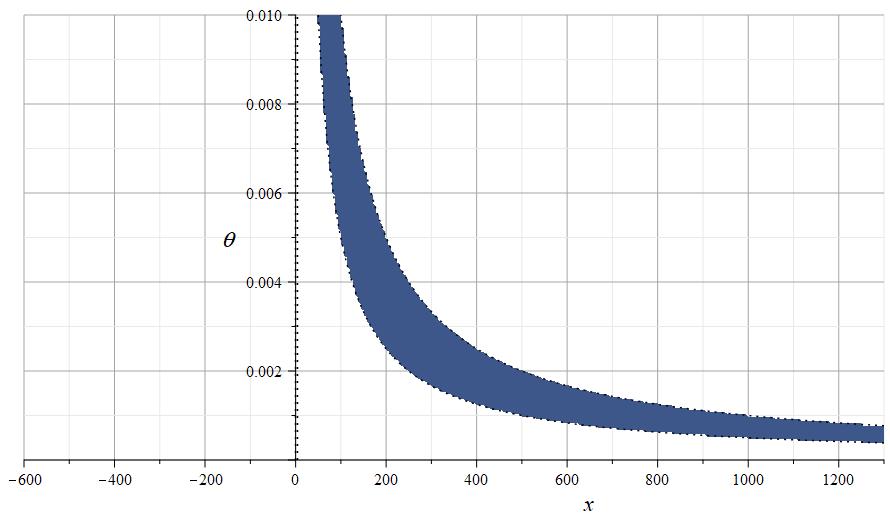}
    \caption{The shaded region indicates the allowed values of $x$ and $\theta$ }
    \label{fig:8}
\end{figure}

\begin{figure}[h!]
    \centering
    \begin{subfigure}{.5\textwidth}
    \centering
    \includegraphics[width=.9\linewidth,height=0.55\linewidth]{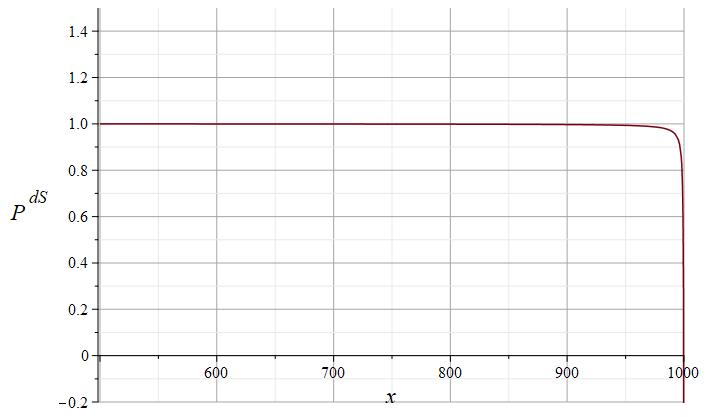}
    \caption{de-Sitter value of $P$ against $x$}
    \end{subfigure}%
    \begin{subfigure}{.5\textwidth}
    \centering
    \includegraphics[width=1\linewidth,height=0.55\linewidth]{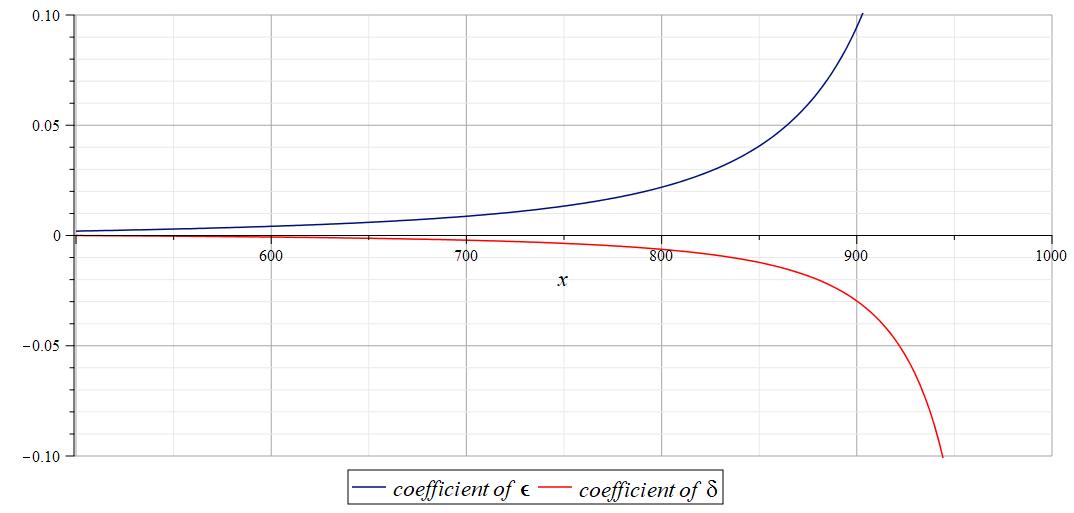}
    \caption{Slow roll coefficients against $x$}
    \end{subfigure}
    \caption{Estimate of the functions involved in the $c_T^2$ expression against x}
    \label{fig:9}
\end{figure}

Here, we analyze the perturbations in the non-minimal model. Initially, while considering only the perturbations in the tensor field $B_{\mu\nu}$, it is shown in Ref. \textcolor{blue}{\cite{Aashish:2019zsy}} that the non-minimal couplings do not contribute to the kinetic terms. Hence, the coefficient matrix of the kinetic terms in Fourier space will be the same as in Eq. (\ref{4.4}), and the results will be identical to the minimal case. Hence, there are no ghost instabilities, given $\tau>0$. Next, we study the tensor perturbations coming from the metric side. The second order part of the perturbed action will look like,
\begin{dmath}\label{6.1}
    S_2=S_2^{Min}+\xi[24(H\phi^2-2\phi\dot{\phi})(h_{\times}\dot{h_{\times}}+h_+\dot{h_+})-6\phi^2(\dot{h_{\times}}^2+\dot{h_{+}}^2)-(h_{\times}^2+h_+^2)(\dot{H}+2H^2)12\phi^2-18\phi^2(h_{\times,z}^2+h_{+,z}^2)-24(h_{\times}h_{\times,zz}+h_{+}h_{+,zz})]+S'_{xy}
\end{dmath}
where, $S_2^{Min}$ is given by Eq. (\ref{4.7}) and $S'_{xy}$ consists of terms involving partial derivatives in $x$ and $y$ coordinates. Moving to Fourier space with the momentum vector $\Vec{k}$ in the $z$ direction, as in the minimal model the perturbed action can be written as a summation of the form Eq. (\ref{4.9}) with the coefficients now modified to,
\begin{equation}\label{6.2}
\begin{array}{c}
    \Omega_k=1+\kappa\tau\phi^2(4-6x), \hspace{1cm} \Omega_c=-2H+2\phi\dot{\phi}\kappa\tau(1-12x)+4\kappa\tau\phi^2H(1+3x) \\
    \vspace{0.1mm}\\
    \Omega_g=\kappa m^2\left(\phi^2+\frac{\phi_0^2}{2} \right)-6(\dot{H}+2H^2)(1+2x\kappa\tau\phi^2)+\kappa\tau(\dot{\phi}+2H\phi)^2-\frac{k^2}{a^2}(1-6x\kappa\tau\phi^2) 
    \end{array}
\end{equation}
The positivity of $\Omega_k$ will ensure the absence of ghosts. In the slow roll limit, we can write,
\begin{equation}\label{6.3}
    \kappa\tau\phi^2=\frac{3-\theta(1+6x)}{6(1-3x)}+\epsilon\left(\frac{6+81x-\theta(1+15x+243x^2)}{18(1-3x)^2} \right)+\delta\left(\frac{3-108x+\theta(-2+33x+216x^2)}{36(1-3x)^2} \right)
\end{equation}
Thus, we can approximate $\Omega_k$ up to first order in slow roll parameters as,
\begin{dmath} \label{6.4}
    \Omega_k\approx\frac{9-18x+\theta(18x^2-9x-2)}{3(1-3x)}+\frac{-2+3x}{9(1-3x)^2}[\epsilon(-6-81x+\theta(243x^2+15x+1))\\-\frac{\delta}{2}(3-108x+\theta(216x^2+33x-2))]
\end{dmath}
In Eq. (\ref{6.4}) there are four parameters, $x, \theta, \epsilon$ and $\delta$. $\epsilon$ is expected to take values between 0 and 1. $\delta$ can be naively related with $\epsilon$ through Eq. (\ref{5.11}). $x$ and $\theta$ values are constrained from the positivity of the de-Sitter solutions in Eq. (\ref{5.5}). The allowed $x$ and $\theta$ values satisfying the condition $\Omega_k>0$ and the previous constraints are depicted in Fig. (\ref{fig:7}) for $\epsilon$ values with different orders.\\
\begin{figure}[h!]
    \centering
    \begin{subfigure}{.5\textwidth}
    \centering
    \includegraphics[width=.9\linewidth,height=.6\linewidth]{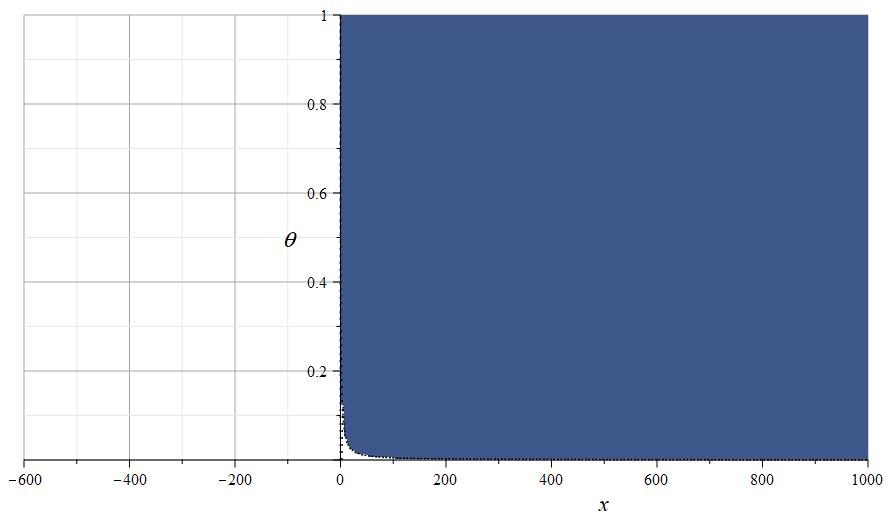}
    \caption{$\epsilon=1$}
    \end{subfigure}%
    \begin{subfigure}{.5\textwidth}
    \centering
    \includegraphics[width=.9\linewidth,height=.6\linewidth]{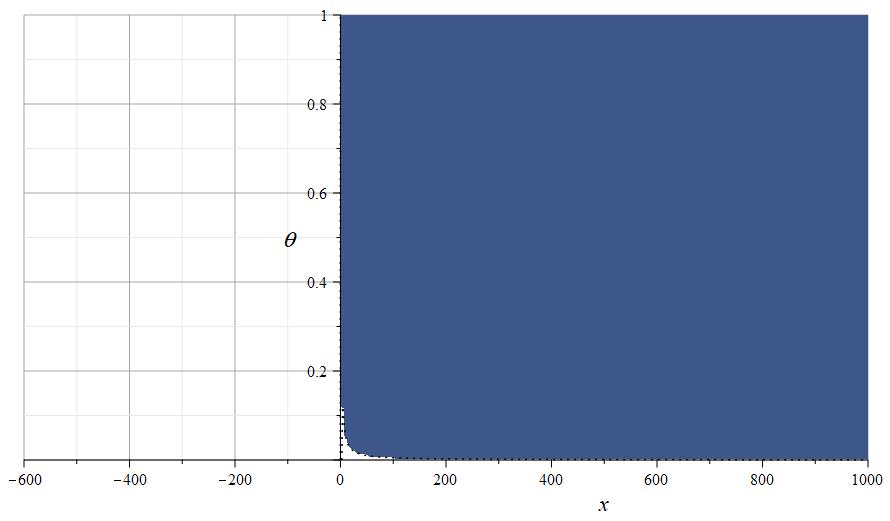}
    \caption{$\epsilon=0.1$}
    \end{subfigure}
\end{figure}
\\
We can see from Fig. (\ref{fig:7}) that as the order of $\epsilon$ decreases the allowed parameter space is becoming smaller. For $\epsilon$ orders smaller than $10^{-3}$, there is no significant change in the available parameter space. The enlarged view of the allowed region is shown in Fig. (\ref{fig:8}).
We can see that the allowed region is particularly broader for $\theta$ values around $10^{-3}$. So, we have chosen a $\theta$ value of about 0.001. For this $\theta$, $x$ can take values from 500 to 1000. Since we have eliminated the possibility of ghost instabilities, we can move on to estimate the primordial GW velocity. In the quasi de-Sitter limit, we have, 

\begin{equation} \label{6.5}
    \frac{\dot{\Omega_k}}{\Omega_k}=\frac{2\kappa\tau\phi^2(4-6x)}{1+\kappa\tau\phi^2(4-6x)}H\delta , \hspace{2em}  \frac{\dot{\Omega_c}+3H\Omega_c-\Omega_g}{\Omega_k}=Q(x)H^2+\frac{k^2}{a^2}P(x)
\end{equation}
where the functions $Q$ and $P$ are given as,
\begin{equation} \label{6.6}
   Q(x)=\frac{8\kappa\tau\phi^2(\delta+2)}{1+\kappa\tau\phi^2(4-6x)}, \hspace{2em} P(x)=1-\frac{4\kappa\tau\phi^2}{1+\kappa\tau\phi^2(4-6x)}
\end{equation}
Following the same recipe we used for the minimal model in Section \ref{sec:4}, in the deep subhorizon limit, we will have $c_T^2=P(x)$. So,
\begin{dmath} \label{6.7}
    c_T^2\approx 1-\frac{6-2\theta(1+6x)}{9-18x-\theta(2-3x)(1+6x)}-\epsilon[\frac{2(6+81x-\theta(1+15x+243x^2))}{(9-18x-\theta(2+9x-18x^2))^2}]+\delta[ \frac{108x-3+\theta(2-33x-216x^2)}{(9-18x-\theta(2+9x-18x^2))^2}]
\end{dmath}
An estimate of these functions in our parameter limit is shown in Fig. (\ref{fig:9}).
\\
We can see that the coefficients of the slow roll parameters in Fig. (\ref{fig:9}) are of the order of  $10^{-2}-10^{-3}$ for the $x$ values less than 900. The values are rising rapidly and get to higher orders as $x$ gets larger. We already selected $x$ to lie between 700 and 1000 from the slow roll analysis. We now limit $x$ values such that $700<x<900$. The slow roll coefficients in Eq. (\ref{6.7}) are now of the order $10^{-3}-10^{-2}$. The smallness of the slow roll parameters further reduce their contributions to the velocity $c_T^2$. Thus $c_T^2$ is dominated by the contribution from its de-Sitter part which almost equals 1 in natural units which can be identified from Fig. (\ref{fig:9}).
Thus, we are able to obtain a primordial gravitational wave velocity equal to $c$ by adding the coupling with Ricci tensor. \\ \\
Now we look at the evolution of the tensor modes from Eq. (\ref{4.13}). We adopt the analysis performed in Refs. \textcolor{blue}{\cite{Dodelson:2003ft}} and \textcolor{blue}{\cite{Aashish:2020mlw}}. The equations from now onwards will be expressed in the conformal time coordinate $\eta$. The derivative with respect to $\eta$ will be represented by a prime over the quantity. Substituting the explicit forms of the coefficients in Eq. (\ref{4.13}), we have,
\begin{equation} \label{6.8}
    h_e''+\left(2+\frac{(4-6x)(3-\theta(1+6x))}{9-18x+\theta(18x^2-9x-2)}\delta \right)aHh_e'+[k^2+a^2H^2Q(x)]h_e=0
\end{equation}
where, we have approximated $P(x)$ as 1. Using the transformation, $h_e=a^{-\lambda}\Tilde{h_e}$, with $\lambda=1+\frac{(2-3x)(3-\theta(1+6x))}{9-18x+\theta(18x^2-9x-2)}\delta$, we can rewrite the above equation as,
\begin{equation} \label{6.9}
    \Tilde{h_e}''+\Tilde{h_e}\left(k^2+a^2H^2\left( 1+\epsilon-3\lambda+Q(x)\right) \right)=0
\end{equation}
In the quasi de-Sitter limit, we have,
\begin{dmath} \label{6.10}
    Q(x)\approx Q^{dS}+Q^{\epsilon}\epsilon+Q^{\delta}\delta
\end{dmath}
where,
\begin{equation} \label{6.11}
    Q^{dS}=\frac{8(3-\theta(1+6x))}{9-18x+\theta(18x^2-9x-2)}, \hspace{2em} Q^{\epsilon}=\frac{8(6+81x-\theta(1+15x+243x^2))}{(9-18x+\theta(18x^2-9x-2))^2} 
\end{equation}
\begin{equation} \label{ex2}
    Q^{\delta}=\frac{4(30-162x+\theta(378x^2-30x-17)+\theta^2(2+21x+36x^2-108x^3))}{(9-18x+\theta(18x^2-9x-2))^2}
\end{equation}
\begin{figure}[h!]
    \centering
    \begin{subfigure}{.5\textwidth}
    \centering
    \includegraphics[width=.9\linewidth,height=0.55\linewidth]{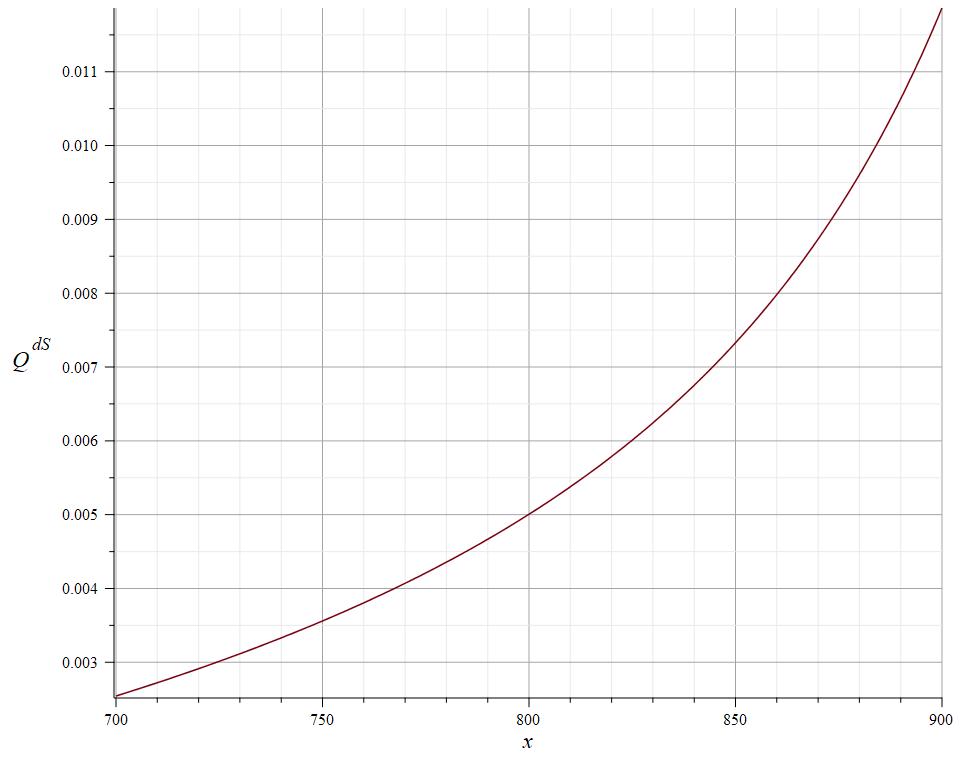}
    \caption{de-Sitter value of $Q$ against $x$}
    \end{subfigure}%
    \begin{subfigure}{.5\textwidth}
    \centering
    \includegraphics[width=.9\linewidth,height=0.55\linewidth]{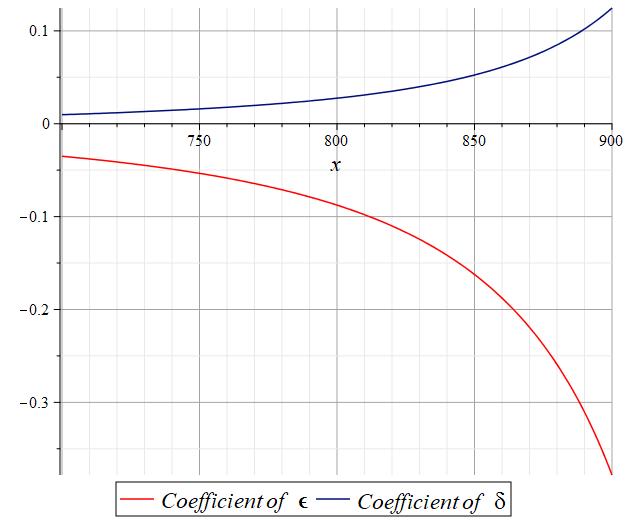}
    \caption{Slow roll coefficients against $x$}
    \end{subfigure}
    \caption{Estimate of the functions involved in the $Q(x)$ expression against x}
    \label{fig:Q}
\end{figure}
\\
An estimate of these functions is performed in Fig. (\ref{fig:Q}). From Fig. (\ref{fig:Q}), we can see that the de-Sitter part of $Q$ is of the order of $10^{-3}-10^{-2}$ whereas, the slow roll coefficients $Q^{\epsilon}$ and $Q^{\delta}$ are roughly of the order of $10^{-2}-10^{-1}$.
Thus,
\begin{equation}\label{6.12}
     \Tilde{h_e}''+\Tilde{h_e}\left(k^2+a^2H^2\left( 1-3\lambda+Q^{dS}+\epsilon\left(1+Q^{\epsilon} \right)+Q^{\delta}\delta\right) \right)=0
\end{equation}
In the slow roll limit, we have, $aH\approx-(1+\epsilon)/\eta$. Applying this, we obtain an equation of the form,
\begin{equation} \label{6.13}
    \Tilde{h_e}''+\left(k^2-\frac{\omega^2}{\eta^2}\right)\Tilde{h_e}=0
\end{equation}
where, 

\begin{equation} \label{6.14}
    \omega^2=3\lambda-1-Q^{dS}+3\epsilon
\end{equation}
Here, we neglected the terms that are smaller in magnitude than the order of the slow roll parameters. Eq. (\ref{6.14}) is the equation of a harmonic oscillator. We now quantize the oscillator. A quantum operator can be decomposed in terms of the creation and annihilation operators as follows.
\begin{equation} \label{6.15}
    \hat{\Tilde{h_e}}(k,\eta)=\nu_e(k,\eta)\hat{a}_{\Vec{k}}+\nu_e(k,\eta)^*\hat{a}^{\dagger}_{\Vec{k}}
\end{equation}
where $\nu_e(k,\eta)$ satisfies the equation,
\begin{equation} \label{6.16}
     \nu_e(k,\eta)''+\left(k^2-\frac{\omega^2}{\eta^2}\right)\nu_e(k,\eta)=0
\end{equation}
For the initial conditions, we assume the universe at early times to be in the Bunch-Davies vacuum state \textcolor{blue}{\cite{Kundu:2011sg}}. For the subhorizon case, $k>>\frac{1}{|\eta|}$, we can neglect the $\frac{\omega^2}{\eta^2}$ term in the bracket. Hence, the properly normalized solution will be,
\begin{equation} \label{6.17}
    \nu_e(k,\eta)=\frac{1}{\sqrt{2k}} e^{-ik\eta}
\end{equation}
which is a plane wave. In the super horizon case, $k<<\frac{1}{|\theta|}$, we can neglect the $k^2$ term in the bracket. So, we get the solutions,
\begin{equation} \label{6.18}
    \nu_e\propto a^{-1-\lambda+\epsilon+\frac{1}{3}Q^{dS}} \hspace{1em} or \hspace{1em} \nu_e\propto a^{\lambda-\frac{1}{3}Q^{dS}} \hspace{1em} \Rightarrow \hspace{1em} h_e\propto a^{-1-2\lambda+\epsilon+\frac{1}{3}Q^{dS}} \hspace{1em} or \hspace{1em} h_e\propto a^{-\frac{1}{3}Q^{dS}}
\end{equation}
We are able to get two solutions for $h_e$, of which the second one in Eq. (\ref{6.18}) is not exactly a constant in time. This solution has a damping caused by the $Q^{dS}$ term. From Fig. (\ref{fig:Q}), we can see that $Q^{dS}$ is of the order of $10^{-3}-10^{-2}$. Hence, the damping due to this term is very small and $h_e$ can be treated nearly as a constant in time. This is in contrast to the results of Ref. \textcolor{blue}{\cite{Aashish:2020mlw}}, where the solution $h_{e}$ was strictly a constant. The difference arises due to the $Q(x)$ term in the expression in Eq. (\ref{6.5}). The predecessor model, \textcolor{blue}{\cite{Aashish:2020mlw}}, didn't have any $\tau$ dependence in the background equations. This is because the contribution to $T_{\mu\nu}$ from the $\tau$ term in the action, $T_{\mu\nu}^{\tau}$, vanishes for the choice of $B_{\mu\nu}$ with only the $B_{ij}$ part as non zero. Hence, $Q(x)=0$, and a constant solution was obtained for the tensor modes in the superhorizon scales. But our scenario is different. We have $\tau$ term in our background equations and hence the solution in our case is not a strict constant. The amplitude is negligible when compared with the wavelength. Thus the wavelength is said to be nearly frozen. The general solution of Eq. (\ref{6.16}) can be calculated by rewriting the differential equation in terms of the new variables $p=-k\eta$ and $\Tilde{\nu_e}=p^{-\frac{1}{2}}\nu_e$.
\begin{equation} \label{6.19}
    p^2\frac{d^2\Tilde{\nu_e}}{dp^2}+p\frac{d\Tilde{\nu_e}}{dp}+(p^2-\nu^2)\Tilde{\nu_e}=0
\end{equation}
where $\nu^2=\omega^2+\frac{1}{4}$. Upto first order in slow roll parameters, we can write,
\begin{equation}\label{6.20}
    \nu\approx\frac{1}{2}+\lambda+\epsilon-\frac{Q^{dS}}{3}
\end{equation}
This is the Bessel's differential equation and its exact solution can be written as a sum of the Hankel functions of first and second kind. Thus,
\begin{equation} \label{6.21}
    \nu_e(p)=\sqrt{p}[A_1H_{\nu}^{(1)}(p)+A_2H_{\nu}^{(2)}(p)]
\end{equation}
where $H_{\nu}^{(1)}(p)$ and $H_{\nu}^{(2)}(p)$ are Hankel functions of the first and second kind respectively. In the asymptotic limit where $p>>1$, i.e the subhorizon limit, the Hankel functions take the form \textcolor{blue}{\cite{Guzzetti:2016mkm}}, 
\begin{equation} \label{6.22}
    H_{\nu}^{(1)}(p)\approx \sqrt{\frac{2}{\pi p}}e^{-\frac{i\pi}{4}(1+2\nu)}e^{ip}
\end{equation}
\begin{equation} \label{6.23}
     H_{\nu}^{(2)}(p)\approx \sqrt{\frac{2}{\pi p}}e^{\frac{i\pi}{4}(1+2\nu)}e^{-ip}
\end{equation}
Matching with the plane wave solution in Eq. (\ref{6.17}), we can put $A_2=0$. Then,
\begin{equation} \label{6.24}                          
    A_1=\frac{1}{2}\sqrt{\frac{\pi}{k}}e^{i\left(\nu+\frac{1}{2}\right)\frac{\pi}{2}}
\end{equation}
Thus the exact solution in the subhorizon limit will be,
\begin{equation} \label{6.25}
    \nu_e(k,\eta)=\frac{\sqrt{\pi}}{2}e^{i\left(\nu+\frac{1}{2}\right)\frac{\pi}{2}}\sqrt{-\eta} H_{\nu}^{(1)}(-k\eta)
\end{equation}
Now, we look at the behaviour of this solution on superhorizon scales. Super horizon modes have great significance in the observational aspect. The scales that grow beyond the horizon will reenter the horizon after the end of inflation, leaving their imprints in CMB. Their observation can shed enough light into early universe physics. In the superhorizon limit, i.e $p<<1$, the Hankel function of first kind takes the following asymptotic form.
\begin{equation} \label{6.26}
    H_{\nu}^{(1)}(p)\approx\sqrt{\frac{2}{\pi}}\frac{\Gamma(\nu)}{\Gamma(3/2)}e^{-i\frac{\pi}{2}}2^{\nu-\frac{3}{2}}p^{-\nu}
\end{equation}
Thus in the super horizon limit we have,
\begin{equation} \label{6.27}
    \nu_e(k,\eta)=\frac{\Gamma(\nu)}{\Gamma(3/2)}e^{i\left(\nu-\frac{1}{2}\right)\frac{\pi}{2}}2^{\nu-\frac{3}{2}}\frac{1}{\sqrt{2k}}(-k\eta)^{\frac{1}{2}-\nu}
\end{equation}
where $\Gamma$ is the Euler function. In the slow roll limit $a$ varies as $\eta^{-(1+\epsilon)}$. So the tensor modes now become,
\begin{equation}\label{6.28}
    h_e(k,\eta)=\frac{\nu_e(k,\eta)}{a}=\frac{C}{k^{\lambda+\epsilon+\frac{1}{2}}}(-k\eta)^{\lambda+\epsilon+\frac{1}{2}-\nu}
\end{equation}
where $C$ is given by $\Gamma(\nu)2^{\nu-1}(-1)^{-\lambda-\epsilon}e^{i(\nu-\frac{1}{2})\frac{\pi}{2}}/\sqrt{\pi}$. Here $\nu$ is as given in Eq. (\ref{6.20}). So,
\begin{equation} \label{6.29}
    h_e(k,\eta)=\frac{C}{k^{\lambda+\epsilon+\frac{1}{2}}}(-k\eta)^{\frac{1}{3}Q^{dS}}
\end{equation}
Thus $h_e$ is almost a constant in proper time on superhorizon scales. But, it is scale dependent since it depends on $k$. We can calculate the power spectrum of tensor perturbations. We have the power spectrum,
\begin{equation} \label{6.30}
    P(k)=\frac{k^3}{2\pi^2}\sum_e|h_e|^2=|A|^2k^{2(1+\frac{2}{3}Q^{dS}-\lambda-\epsilon)}
\end{equation}
where $|A|^2$ is a constant which has an extremely small time dependence. The spectral index can be calculated as,
\begin{equation} \label{6.31}
    n=\frac{dlnP}{dlnk}=-2\epsilon-\frac{(4-6x)(3-\theta(1+6x))}{9-18x+\theta(18x^2-9x-2)}\delta+\frac{16(3-\theta(1+6x))}{3(9-18x+\theta(18x^2-9x-2))}
\end{equation}
Thus the power spectrum is not exactly independent of scales, but is nearly scale invariant. The spectral index in typical scalar field models depends only on the slow roll parameter $\epsilon$. The predecessor antisymmetric tensor field model in Ref. \textcolor{blue}{\cite{Aashish:2020mlw}} had the spectral index depending on both the slow roll parameters $\epsilon$ and $\delta$. But, here we have a slight variance even at the de-Sitter level, whereas the general inflationary models predict a scale invariant power spectrum in the de-Sitter limit. This variance arises due to the presence of the function $Q(x)$. Here, the spectral index has a value of the order of $10^{-3}$ at the de-Sitter limit, whereas additional dependencies on both the slow roll parameters through Eq. (\ref{6.31}) arise in the slow roll case unlike \textcolor{blue}{\cite{Aashish:2020mlw}}.

\section{The Quartic potential} \label{sec:7}
\begin{figure}[h!]
    \centering
    \includegraphics[width=.7\linewidth]{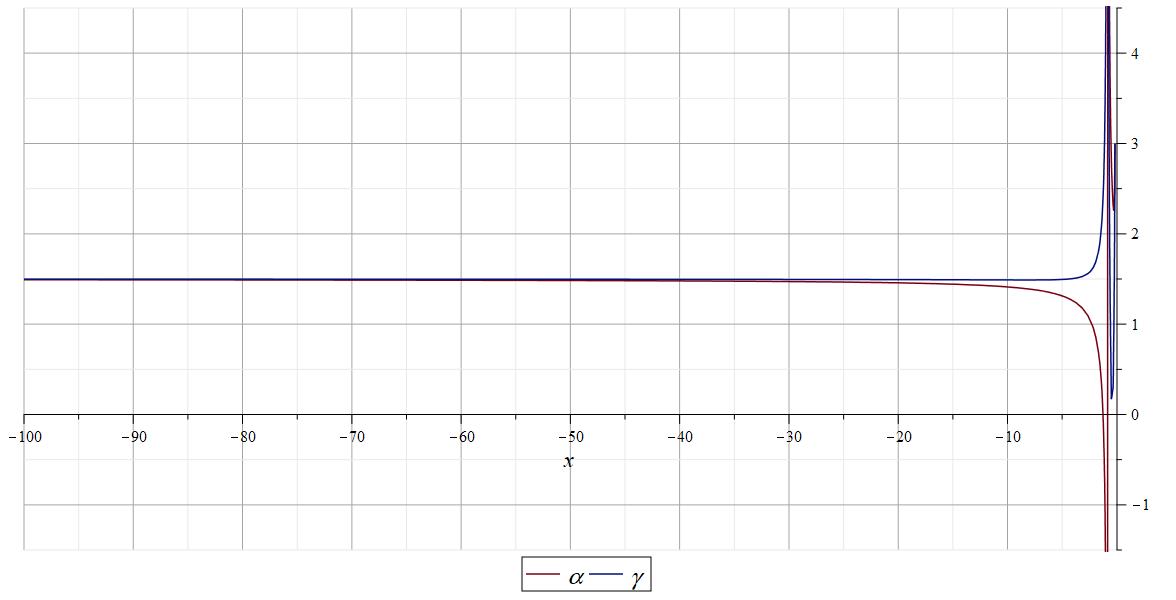}
    \caption{$\alpha$ and $\gamma$ values against $x$}
    \label{fig:10}
\end{figure}

\begin{figure}[h!]
    \centering
    \includegraphics[width=.7\linewidth]{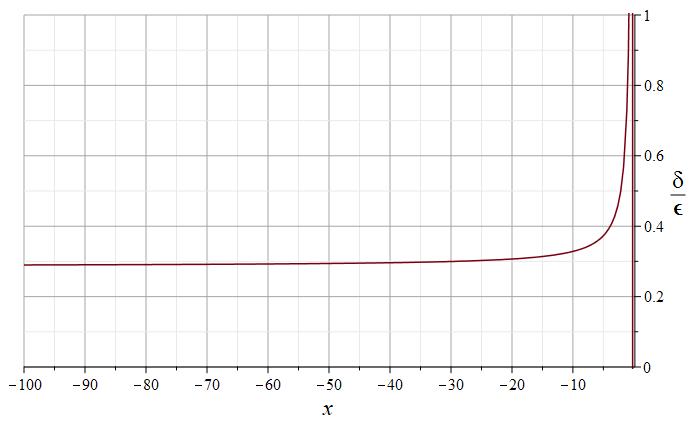}
    \caption{Ratio of slow roll parameters against $x$}
    \label{fig:11}
\end{figure}

\begin{figure}[h!]
    \centering
    \includegraphics[width=.7\linewidth,height=.4\linewidth]{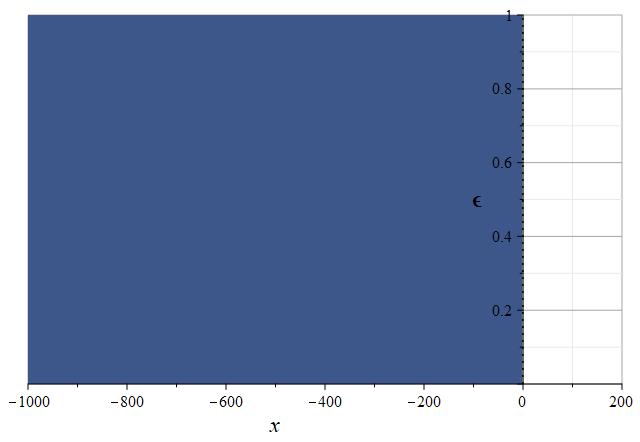}
    \caption{The shaded region indicates the region where $\Omega_k$ is positive}
    \label{fig:12}
\end{figure}

\begin{figure}[h!]
    \centering
    \begin{subfigure}{.5\textwidth}
    \centering
    \includegraphics[width=1\linewidth,height=.6\linewidth]{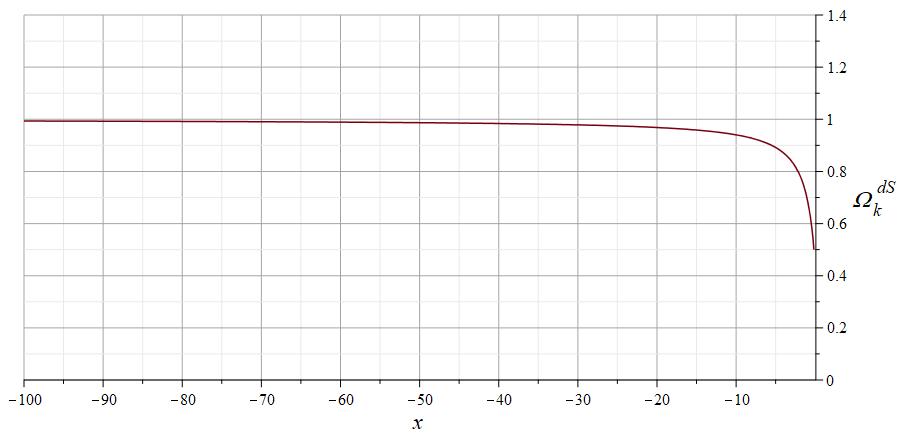}
    \caption{de-Sitter value of $P$ against $x$}
    \end{subfigure}%
    \begin{subfigure}{.5\textwidth}
    \centering
    \includegraphics[width=1\linewidth,height=.6\linewidth]{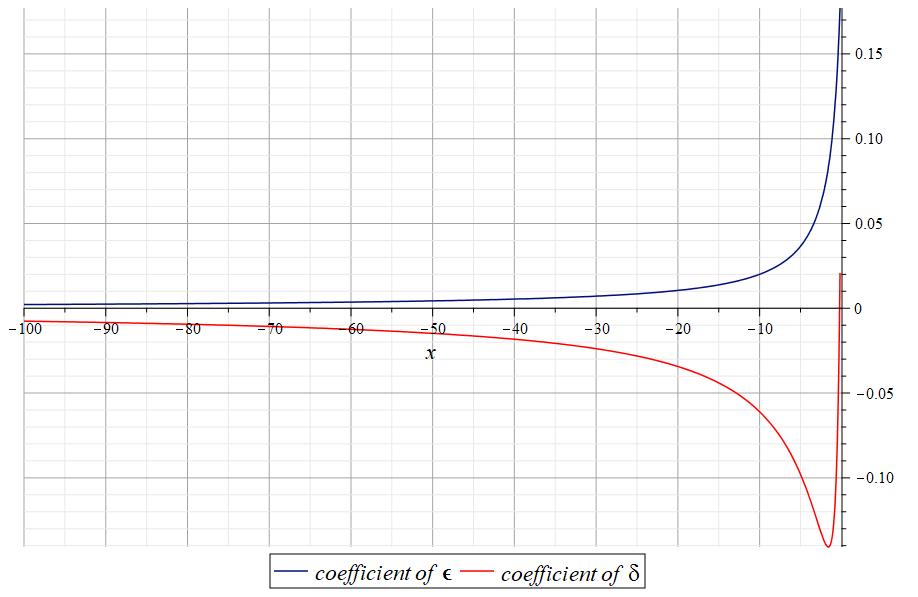}
    \caption{Slow roll coefficients against $x$}
    \end{subfigure}
    \caption{Estimate of the functions involved in the $c_T^2$ expression against x}
    \label{fig:13}
\end{figure}

\begin{figure}[h!]
    \centering
    \includegraphics[width=.65\linewidth,height=.38\linewidth]{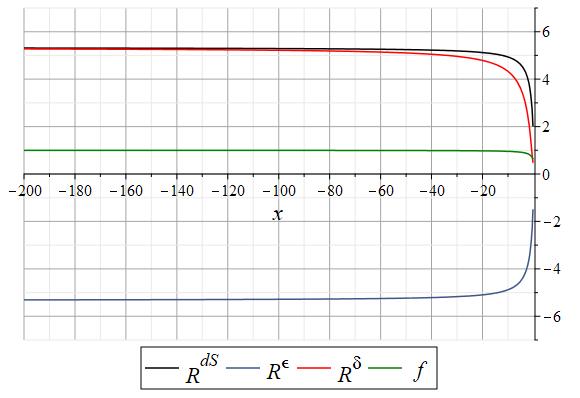}
    \caption{Estimate of the functions involved in the expression for $\omega^2$}
    \label{fig:15}
\end{figure}

\begin{figure}[h!]
    \centering
    \includegraphics[scale=0.3]{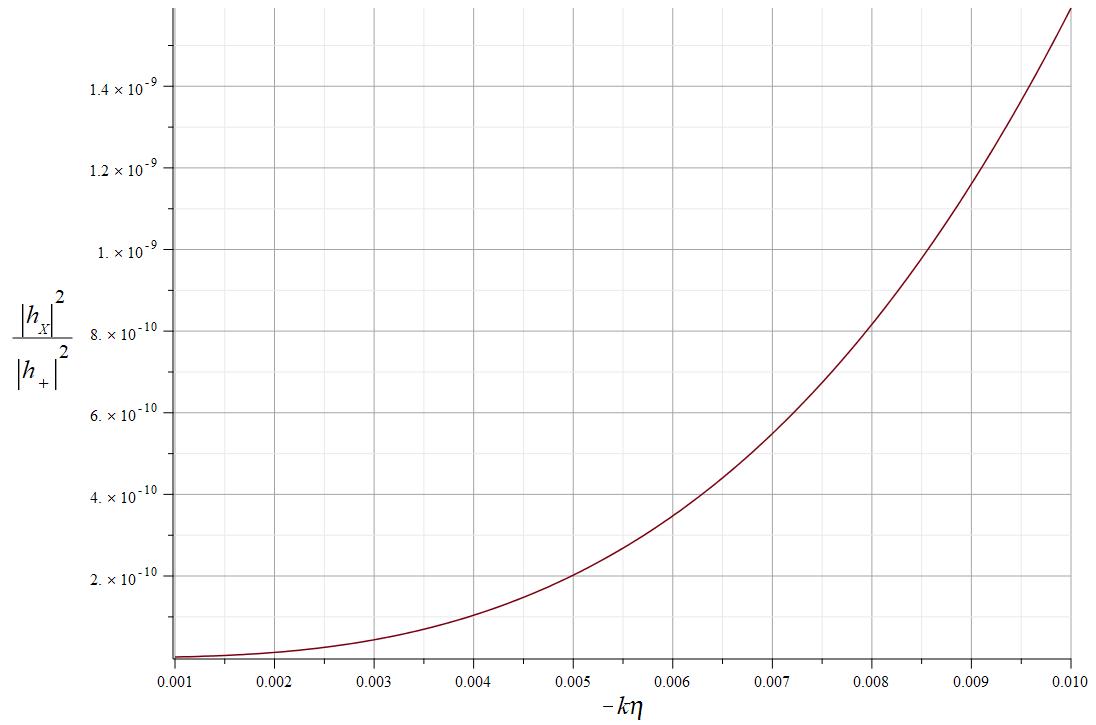}
    \caption{Plot of $\frac{|h_{\times}|^2}{|h_+|^2}$ against $-k\eta$}
    \label{fig:16}
\end{figure}

In section \ref{sec:3}, we talked about the limitation posed by the quadratic potential. For a smooth transition to the subsequent reheating phase after the inflationary era, there should be a minimum in the potential, which our quadratic potential in Eq. (\ref{2.6}) misses. In order to resolve this, we can modify our initial potential to a quartic one in $B_{\mu\nu}$. Quartic potentials have been studied rigorously in the context of scalar field inflation \textcolor{blue}{\cite{Pozdeeva:2020apf,Kannike:2015kda,Bostan:2019fvk}}. In our case, we have,
\begin{equation} \label{7.1}
    V=\lambda(B_{\mu\nu}B^{\mu\nu})^2=36\lambda\phi^4
\end{equation}
where $\lambda$ is a positive dimensionless constant.
We can see from Eq. (\ref{7.1}) that the potential varies with $\phi$ as $\phi^4$. Thus a minimum in the driving potential $V(\phi)$ can be ensured making a smooth exit to the reheating phase possible. In this case, $T_{\mu\nu}^M$ given in Eq. (\ref{3.3}) now gets modified to,
\begin{equation} \label{7.2}
    T_{\mu\nu}^M=\frac{1}{2}{H^{\alpha\beta}}_{\mu}H_{\nu\alpha\beta}+8\lambda{B^{\alpha}}_{\mu}B_{\alpha\nu}B_{\rho\sigma}B^{\rho\sigma}-g_{\mu\nu}\left(\frac{1}{12}H_{\alpha\beta\gamma}H^{\alpha\beta\gamma} +\lambda (B_{\alpha\beta}B^{\alpha\beta})(B_{\rho\sigma}B^{\rho\sigma})\right)
\end{equation}
We keep the same background structure of $B_{\mu\nu}$ as in Eq. (\ref{2.4}). After imposing the homogeneity and isotropy conditions in the energy momentum tensor, the system of equations for this new potential will be,
\begin{equation} \label{7.3}
    H^{2}=-60\kappa\lambda\phi^4-\kappa\tau(H\phi\dot{\phi}+2H^{2}\phi^{2})
\end{equation}
\begin{equation} \label{7.4}
    2\dot{H}+3H^{2}=-36\kappa\lambda\phi^4+3\kappa\tau(H\phi\dot{\phi}+2H^{2}\phi^{2})
\end{equation}
 Applying the de-Sitter conditions in these equations, we get the relation,
 \begin{equation} \label{7.5}
    \frac{H_d^2}{\phi_d^2}=-\frac{12\lambda}{\tau}
\end{equation}
For the quadratic potential, Eq. (\ref{2.6}), the ghost free conditions for the perturbed $B_{\mu\nu}$ modes is $\tau>0$. Since in the quartic model, Eq. (\ref{7.1}), the potential doesn't contain any time derivatives similar to the quadratic model, here too the ghost free conditions will be $\tau>0$. Thus from Eq. (\ref{7.5}), $H_d^2$ and $\phi_d^2$ cannot be simultaneously made positive. So, we need to add non-minimal couplings. Analogous to the previous case, we can add the  $RB_{\mu\nu}B^{\mu\nu}$ coupling parameterised by the parameter $\xi$. Now the Einstein equations become,
\begin{equation} \label{7.6}
    H^{2}=-60\kappa\lambda\phi^4-\kappa\tau(H\phi\dot{\phi}+2H^{2}\phi^{2})-6\xi(3\dot{H}\phi^2-2H\phi\dot{\phi}+5H^2\phi^2)
\end{equation}
\begin{equation} \label{7.7}
    2\dot{H}+3H^{2}=-36\kappa\lambda\phi^4+3\kappa\tau(H\phi\dot{\phi}+2H^{2}\phi^{2})+6\xi(2\phi\ddot{\phi}+2\dot{\phi}^2+4H\phi\dot{\phi}-\dot{H}\phi^2-3H^2\phi^2)
\end{equation}
The condition in Eq. (\ref{7.5}) now becomes,
\begin{equation} \label{7.8}
    \frac{H_d^2}{\phi_d^2}=-\frac{12\lambda}{\tau(1+6x)}
\end{equation}
where $x$ is $\frac{\xi}{\kappa\tau}$. By constraining $x$ to be less than $-\frac{1}{6}$, we can make this ratio positive.
From the equations we can see that the solutions for $\phi_d^2$ are,
\begin{equation} \label{7.9}
    \phi_d^2=\frac{1}{3\kappa\tau} \hspace{0.5cm} or \hspace{0.5cm} 0
\end{equation}
We select the non zero solution. Now our de-Sitter solutions are given by,
\begin{equation} \label{7.10}
    \phi_d^2=\frac{1}{3\kappa\tau}, \hspace{1cm} H_d^2=-\frac{4\lambda}{\kappa\tau^2(1+6x)}
\end{equation}
Now, we look for the stability around this de-Sitter background. As we did for the earlier case, we can write the fluctuation $\Xi$ as,
\begin{equation} \label{7.11}
    \Xi=e^{-\alpha H_dt}(Ce^{\gamma  H_dt}+De^{-\gamma H_dt})
\end{equation}
with $\gamma=\sqrt{\alpha^2-\beta}$. The functions $\alpha(x)$ and $\beta(x)$ are given as,
\begin{equation} \label{7.12}
    \alpha(x)=\frac{5+38x+300x^2+216x^3}{-1-2x+120x^2+144x^3}, \hspace{2em} \beta(x)=\frac{24(12x^2-4x-1)}{144x^3+120x^2-2x-1}
\end{equation}
 The values of $\alpha(x)$ and $\gamma(x)$ for our given range is depicted in the Fig. (\ref{fig:10}). We can see that as $x<-30$, then $\alpha\approx\gamma$. So, in this limit,
\begin{equation} \label{7.13}
    \Xi\approx C+De^{-2\alpha H_dt}
\end{equation}
$\alpha$ is positive in our given range. If $C\approx0$, then we can see that the perturbations decay in time, and we have a stable solution. Even though how such a solution can be made possible is still unknown, we can argue for the vanishing of $C$, and we leave the underlying reasons for future discussion.
\\ \\
Now, using the Eqs. (\ref{7.6}) and (\ref{7.7}), we can obtain a linear relation between the slow roll parameters $\epsilon$ and $\delta$ as follows,
\begin{equation} \label{7.14}
    \delta=\epsilon\left(\frac{-30+24x}{18+84x}\right)
\end{equation}
A plot of $\delta/\epsilon$ against the allowed range of $x$ is given in Fig. (\ref{fig:11}). We can see that towards the negative region the ratio is attaining almost a constant value. Also, the slow roll parameters are roughly of the same order. Following the same arguments we posed for quadratic case, here too we can ensure enough number of e-folds.
 Then we look at perturbations coming solely from $B_{\mu\nu}$. Since, this new potential doesn't have any derivative terms in $B_{\mu\nu}$, the kinetic terms of the perturbed modes in Fourier space won't be different from those calculated in the quadratic model. Hence, ghost instabilities won't be present in the limit $\tau>0$. Then, we study the tensor perturbations coming from the metric side. The second order part of the perturbed action reads,
\begin{dmath} \label{7.15}
S_2=\int dtd^3x \frac{a^3}{4\kappa}[(\dot{h_{\times}}^2+\dot{h_{+}}^2)(1+4\kappa\tau\phi^2-6\xi\phi^2)+(h_{\times}\dot{h_{\times}}+h_+\dot{h_+})(4\kappa\tau(\phi\dot{\phi}+2H\phi^2)-4H\\+24\xi (H\phi^2-2\phi\dot{\phi}))+(h_{\times}^2+h_+^2)(\kappa\tau(\dot{\phi}+2H\phi)^2-6(\dot{H}+      2H^2)(1+2\xi))+(4(h_{\times}h_{\times,zz}+h_{+}h_{+,zz})+3(h_{\times,z}^2+h_{+,z}^2))(1-6\xi\phi^2)-\kappa\lambda\phi^4(184h_{\times}^2+120h_+^2)]+S_{xy}
\end{dmath}
$S_{xy}$ in Eq. (\ref{7.15}) consists of terms involving partial derivatives of the tensor modes with respect to the position coordinates-$x$ and $y$, which will vanish once we go to Fourier space with the momentum vector $\Vec{k}$ lying along the $Z$ direction. Here too, the tensor modes $h_{+}$ and $h_{\times}$ decouple, and the action now reads,
\begin{equation} \label{7.16}
    S_2^{FT}=\sum_{e=+,\times}\int dtd^3k\frac{a^3}{4\kappa}[\Omega_k^e\dot{h_e}^{\dagger}\dot{h_e}+\Omega_c^e(\dot{h_e}^{\dagger}h_e+h_e^{\dagger}\dot{h_e})+\Omega_g^eh_e^{\dagger}h_e]
\end{equation}
\begin{equation}\label{7.17}
\begin{array}{c}
    \Omega^{\times}_k=\Omega^+_k=1+\kappa\tau\phi^2(4-6x), \hspace{1em} \Omega^{\times}_c=\Omega^+_c=-2H+2\kappa\tau\phi\dot{\phi}(1-12x)+4H\kappa\tau\phi^2(1+3x) \\
    \vspace{0.1mm}\\
     \Omega_g^+=-120\kappa\lambda\phi^4-6(\dot{H}+2H^2)(1+2x\kappa\tau\phi^2)+\kappa\tau(\dot{\phi}+2H\phi)^2-\frac{k^2}{a^2}(1-6x\kappa\tau\phi^2) \\
     \vspace{0.1mm}\\
     \Omega_g^{\times}=-184\kappa\lambda\phi^4-6(\dot{H}+2H^2)(1+2x\kappa\tau\phi^2)+\kappa\tau(\dot{\phi}+2H\phi)^2-\frac{k^2}{a^2}(1-6x\kappa\tau\phi^2)
    \end{array}
\end{equation}
We can see that, unlike the quadratic case, here the coefficient $\Omega_g$ is different for the two modes. The difference occurs due to the $\lambda$ term. For the quadratic potential, both the polarizations had the same coefficients in the second order of perturbations as in Eq. (\ref{6.2}), but there was an asymmetry at the linear order perturbations. In the quartic potential case, the linear order perturbations in $g_{\mu\nu}$ become second order creating the aforementioned asymmetry. This difference will have its implications and lead to phenomenology different from the model with quadratic potential. The polarizations will have different evolution equations and we may be able to distinguish them from their evolutionary tracks.
Now, for our theory to be devoid of ghosts, the coefficient of the kinetic terms, $\Omega_k^e$ must be positive. From the field equations, we can write,
\begin{equation} \label{7.18}
    \kappa\tau\phi^2\approx\frac{1}{3}+\frac{4x-5}{18}\epsilon-\frac{3+14x}{18}\delta+O(\epsilon^2,\delta^2,\epsilon\delta)
\end{equation}
Thus, we have,
\begin{equation} \label{7.19}
    \Omega_k^e\approx\frac{7}{3}-2x+\frac{-12x^2+23x-10}{9}\epsilon+\frac{42x^2-19x-6}{9}\delta \hspace{2em} (e=+,\times)
\end{equation}
The slow roll parameter $\epsilon$ can vary between 0 and 1. Also, $x$ is constrained to be less than -1/6. $\delta$ can be crudely related with $\epsilon$ through the relation in Eq. (\ref{7.14}). The region satisfying the positivity of $\Omega_k$ for different $x$ and $\epsilon$ values is shown in Fig. (\ref{fig:12}).
\\
Here, we have the freedom to select a wide range of $x$ values. Now, we look at the equation of motion of the tensor modes. The equation of motion, Eq. (\ref{4.13}), will be different for each of the modes, since $\Omega_g$ is different in each case. First we look at the $+$ mode. In this case, we can write,
\begin{equation} \label{7.20}
    \frac{\dot{\Omega_k^+}}{\Omega_k^+}\approx\frac{8-12x}{7-6x}H\delta, \hspace{3em} \frac{\dot{\Omega_c^+}+3H\Omega_c^+-\Omega_g^+}{\Omega_k^+}= Q(x)H^2+\frac{k^2}{a^2}P(x) 
\end{equation}
where $Q(x)$ and $P(x)$ are the same functions we defined in Eq. (\ref{6.6}). As we did for the quadratic case, in the deep subhorizon limit, we can see that
\begin{equation} \label{7.21}
    c_T^2=P(x)\approx \frac{-3+6x}{-7+6x}+\epsilon\left(\frac{-144x+180}{18(-7+6x)^2} \right)+\delta\left(\frac{28x+6}{(-7+6x)^2} \right)
\end{equation}
An estimate of the functions in this equation for our parameter range  is given in Fig. (\ref{fig:13}),\\
\\
We can see that the coefficients of the slow roll parameters are of the order $10^{-2}$ for the range $x<-30$ which we selected from the de-Sitter stability analysis. The smallness of the slow roll parameters further reduces their contribution and hence the value of $c_T^2$ will be dominated by the de-Sitter part. From the figure, we can see that the value becomes closer to 1 when $x$ becomes more negative. Now, for the $\times$ mode,
\begin{equation} \label{7.22}
   \frac{\dot{\Omega_k^{\times}}}{\Omega_k^{\times}}\approx\frac{8-12x}{7-6x}H\delta, \hspace{3em} \frac{\dot{\Omega_c^{\times}}+3H\Omega_c^{\times}-\Omega_g^{\times}}{\Omega_k^{\times}}= R(x)H^2+\frac{k^2}{a^2}P(x)
\end{equation}
where,
\begin{equation} \label{7.23}
    R(x)=\frac{\frac{64\lambda\kappa\phi^4}{H^2}+8\kappa\tau\phi^2(\delta+2)}{1+\kappa\tau\phi^2(4-6x)}\approx{\frac {32-96x}{21-18x}}-\epsilon\left(\frac{576x^2-800x+136}{3(-7+6x)^2} \right)+\delta\left({\frac {576\,{x}^{2}-64\,x+64}{3\, \left( -7+6\,x \right) ^{2}}} 
 \right)
\end{equation}
Here too, in the deep subhorizon limit, we will have, $ c_T^2\approx P(x)$. Hence, both of the tensor modes propagate with the velocity of light in vacuum in the deep subhorizon limit. Now, we can look at the evolution of these tensor modes. Like we did for the quadratic case, we use the conformal time coordinate $\eta$, and employ the transformation $h_e=a^{-\lambda}\Tilde{h_e}$, with $\lambda=1+\frac{4-6x}{7-6x}\delta$. For the $+$ polarization, we have,
\begin{equation} \label{7.24}
    \Tilde{h_+}''+\Tilde{h_+}\left(k^2+a^2H^2\left( 1+\epsilon-3\lambda+Q(x)\right) \right)=0
\end{equation}
Approximating $Q(x)$ to linear order in the slow roll parameters, we can write,
\begin{equation}\label{7.25}
Q(x)\approx \frac{16}{7-6x}+\epsilon\left(\frac{32x-40}{(-7+6x)^2} \right)+\delta\left(\frac{32-160x}{(-7+6x)^2} \right)    
\end{equation}
\begin{figure}[h!]
    \centering
    \begin{subfigure}{.5\textwidth}
    \centering
    \includegraphics[width=1\linewidth]{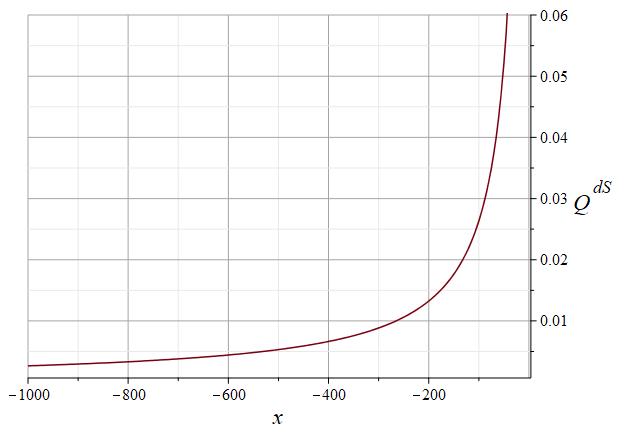}
    \caption{de-Sitter value of $Q$ against $x$}
    \end{subfigure}%
    \begin{subfigure}{.5\textwidth}
    \centering
    \includegraphics[width=1\linewidth]{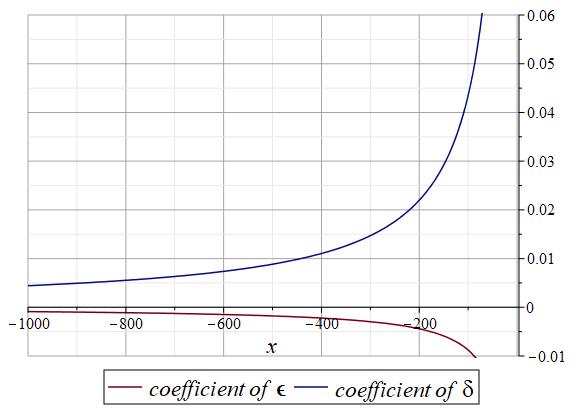}
    \caption{Slow roll coefficients against $x$}
    \end{subfigure}
    \caption{Estimate of the functions involved in the $Q(x)$ expression against x}
    \label{fig:14}
\end{figure}
\\
From Fig. (\ref{fig:14}), it is clear that the de-Sitter part and the coefficients of the slow roll parameters $Q^{\epsilon}$ and $Q^{\delta}$ are all of the order of $10^{-2}-10^{-3}$.
We intend to keep terms only up to the order of slow roll parameters in our equations. So, we can neglect the products of $Q^{\epsilon}$ and $Q^{\delta}$ with the slow roll parameters. Now we have,
\begin{equation} \label{7.26}
    \Tilde{h_+}''+\Tilde{h_+}\left(k^2+a^2H^2\left( 1-3\lambda+\frac{16}{7-6x}+\epsilon\right) \right)=0
\end{equation}
We can convert this equation to the harmonic oscillator equation given in Eq. (\ref{6.13}) with the same form for $\omega^2$.
We quantize the harmonic oscillator following the same procedure we did in section \ref{sec:6}. The solutions will be the same as in the previous case, plane wave solution in the subhorizon case, Eq. (\ref{6.25}), and the nearly frozen solution in the superhorizon limit for $h_{+}$, Eq. (\ref{6.28}). Following the similar steps we did for the quadratic model, we can see that the general solution on superhorizon scales has scale dependence. Thus, the power spectrum will be,
\begin{equation} \label{7.27}
    P_+(k)=|A_+|^2k^{2(1+\frac{2}{3}Q^{dS}-\lambda-\epsilon)}=|A_+|^2k^{2(1+\frac{32}{21-18x}-\lambda-\epsilon)}
\end{equation}
and the spectral index is,
\begin{equation} \label{7.28}
    n_+=-2\epsilon-\frac{4-6x}{7-6x}\delta+\frac{32}{21-18x}
\end{equation}
which suggests a nearly scale invariant spectrum.
For the $\times$ polarization, the $Q(x)$ in Eq. (\ref{7.24}) should be replaced by $R(x)$. Now,
\begin{equation}\label{7.29}
    \omega^2=2-R^{dS}+\epsilon(3-R^{\epsilon}-2R^{dS})+\delta(3f-R^{\delta})
\end{equation}
where $R^{dS}, R^{\epsilon}$ and $R^{\delta}$ are the respective de-Sitter part, coefficient of $\epsilon$ and coefficient of $\delta$ in the expression for $R(x)$ given in Eq. (\ref{7.23}). The function $f$ is defined as $f=\frac{4-6x}{7-6x}$. An estimate of these functions is shown in Fig. (\ref{fig:15}),\\
\\
$\epsilon$ and $\delta$ are very small quantities. So the value of $\omega^2$ will be dominated by the value of $2-R^{dS}$. Thus for the $\times$ mode, $\omega^2$ is a negative quantity. We can define $\omega_1^2=-\omega^2$, so that $\omega=\pm i\omega_1$. In the subhorizon limit, we have the plane wave solution $e^{-ik\eta}/\sqrt{2k}$ for the quantized modes compatible with the Bunch-Davies vacuum condition. In the superhorizon limit, the solution for the quantized mode $\nu_{\times}$ will be,
\begin{equation} \label{7.30}
    \nu_{\times}=C_1a^{(\epsilon-1)(\frac{1}{2}+i\mu)}+C_2a^{(\epsilon-1)(\frac{1}{2}-i\mu)}=a^{-\frac{1}{2}(1-\epsilon)}(C_1e^{-i\mu(1-\epsilon)}+C_2e^{i\mu(1-\epsilon)})
\end{equation}
where $\mu=\frac{\sqrt{4\omega_1^2-1}}{2}\approx1.7$ for our parameter range. Thus the $\times$ mode, on superhorizon scales, will have an oscillatory behaviour that decays in time. The general solution can be calculated by introducing the new variables $p=-k\eta$ and $\Tilde{\nu_{\times}}=p^{-\frac{1}{2}}\nu_{\times}$. We get the Bessel differential equation given in Eq. (\ref{6.19}) but with a different order parameter $\nu$. The order parameter can be written as,
\begin{equation} \label{7.31}
    \nu^2=\omega^2+\frac{1}{4}\hspace{1em}\Longrightarrow\hspace{1em} \nu=\pm i\mu
\end{equation}
Since the order parameter is purely imaginary, a pair of Bessel functions or a pair of Hankel functions do not form a set of satisfactory solutions for the Bessel equation. This is because these functions become imaginary for a purely imaginary order. We can construct suitable linear combinations of Hankel functions which can attain real values in our domain. We use the following functions discussed in Ref. \textcolor{blue}{\cite{dunster90}},
\begin{equation} \label{7.32}
    F_{\rho}(z)=\frac{1}{2}\left(e^{\frac{i\rho\pi}{2}}H_{\rho}^{(1)}(z)+e^{-\frac{i\rho\pi}{2}}H_{\rho}^{(2)}(z) \right)
\end{equation}
\begin{equation} \label{7.33}
    G_{\rho}(z)=\frac{1}{2i}\left(e^{\frac{i\rho\pi}{2}}H_{\rho}^{(1)}(z)-e^{-\frac{i\rho\pi}{2}}H_{\rho}^{(2)}(z) \right)
\end{equation}
Here $\rho$ can be any complex number, even a purely imaginary one. Thus, we can write our general solution as,
\begin{equation} \label{7.34}
    \nu_{\times}=\sqrt{p}(C_1F_{i\mu}+C_2G_{i\mu})
\end{equation}
In the deep subhorizon limit, $p>>1$, these functions take the following form,
\begin{equation}\label{7.35}
    F_{i\mu}(p>>1)\sim \sqrt{\frac{2}{\pi p}}\left(\cos(p-\pi/4)\sum_{s=0}^{\infty}(-1)^s\frac{A_{2s}(i\mu)}{p^{2s}})-\sin(p-\pi/4)\sum_{s=0}^{\infty}(-1)^s\frac{A_{2s+1}(i\mu)}{p^{2s+1}}) \right)
\end{equation}
\begin{equation} \label{7.36}
    G_{i\mu}(p>>1)\sim \sqrt{\frac{2}{\pi p}}\left(\sin(p-\pi/4)\sum_{s=0}^{\infty}(-1)^s\frac{A_{2s}(i\mu)}{p^{2s}})+\cos(p-\pi/4)\sum_{s=0}^{\infty}(-1)^s\frac{A_{2s+1}(i\mu)}{p^{2s+1}}) \right)
\end{equation}
where the coefficients $A_s$ are given by,
\begin{equation} \label{7.37}
    A_s(\lambda)=\frac{(4\lambda^2-1^2)(4\lambda^2-3^2).............(4\lambda^2-(2s-1)^2)}{s!8^s}
\end{equation}
$A_s(i\mu)$ is a real function since it depends only on the square of the argument. We can rewrite the above expressions in Eqs. (\ref{7.35}) and (\ref{7.36})as,
\begin{equation}\label{7.38}
    F_{i\mu}\sim \sqrt{\frac{2}{\pi p}}\left(e^{ip}M(p,\mu)+e^{-ip}M^*(p,\mu) \right)
\end{equation}
\begin{equation}\label{7.39}
    G_{i\mu}\sim \frac{1}{i}\sqrt{\frac{2}{\pi p}}\left(e^{ip}M(p,\mu)-e^{-ip}M^*(p,\mu) \right)
\end{equation}
where $M^*$ is the complex conjugate of $M$. $M$ is given as,
\begin{equation} \label{7.40}
    M(p,\mu)=\frac{e^{-i\pi/4}}{2}\left(\sum_{s=0}^{\infty}(-1)^s\frac{A_{2s}(i\mu)}{p^{2s}}+i\sum_{s=0}^{\infty}(-1)^s\frac{A_{2s+1}(i\mu)}{p^{2s+1}} \right)
\end{equation}
Since $p$ is very large, we can approximate $M$ as,
\begin{equation}\label{7.41}
    M\sim \frac{e^{-i\pi/4}A_0}{2}=-\frac{e^{-i\pi/4}(1+4\mu^2)}{2}
\end{equation}
Thus, in the subhorizon limit, we have,
\begin{equation}\label{7.42}
    \nu_{\times}=\sqrt{\frac{2}{\pi}}(e^{ip}M(C_1-iC_2)+e^{-ip}M^*(C_1+iC_2))
\end{equation}
Matching Eq. (\ref{7.42}) with the plane wave solution $e^{ip}/\sqrt{2k}$ given by the Bunch-Davies vacuum, we get,
\begin{equation}\label{7.43}
    C_1=\frac{1}{4M}\sqrt{\frac{\pi}{k}}, \hspace{2em} C_2=\frac{i}{4M}\sqrt{\frac{\pi}{k}}
\end{equation}
Thus our general solution will look like,
\begin{equation}\label{7.44}
    \nu_{\times}=\frac{1}{4M}\sqrt{\frac{p\pi}{k}}(F_{i\mu}+iG_{i\mu})
\end{equation}
In the superhorizon limit, $p<<1$, 
\begin{equation}\label{7.45}
    F_{i\mu}(p)\sim\sqrt{\frac{2\tanh(\mu\pi/2)}{\mu\pi}}\cos(\mu\ln(p/2)-\phi_{\mu,0})
\end{equation}
\begin{equation}\label{7.46}
    G_{i\mu}(p)\sim\sqrt{\frac{2\coth(\mu\pi/2)}{\mu\pi}}\sin(\mu\ln(p/2)-\phi_{\mu,0})
\end{equation}
where $\phi_{\lambda,s}$ is defined as $\phi_{\lambda,s}=\arg(\Gamma(1+s+i\lambda)) $. In our case $ \phi_{\mu,0}\approx-0.02757$. Thus our solution in the superhorizon limit will look like,
\begin{dmath}\label{7.47}
    h_{\times}=(-1)^{-(\lambda+\epsilon)}\frac{\sqrt{\pi}}{4Mk^{\lambda+\epsilon+\frac{1}{2}}}p^{\lambda+\epsilon+\frac{1}{2}}[\sqrt{\frac{2\tanh(\mu\pi/2)}{\mu\pi}}\cos(\mu\ln(p/2)-\phi_{\mu,0})\\\hspace{10em}+i\sqrt{\frac{2\coth(\mu\pi/2)}{\mu\pi}}\sin(\mu\ln(p/2)-\phi_{\mu,0})]
\end{dmath}
Also, we get,
\begin{dmath}\label{7.48}
    |h_{\times}|^2=\frac{1}{4\mu(1+4\mu^2)^2}\left(\frac{p}{k} \right)^{2\lambda+2\epsilon+1}(A+B\cos(2\mu\ln(p/2)-2\phi_{\mu,0}))
\end{dmath}
where the coefficients $A$ and $B$ are given as,
\begin{equation}\label{7.49}
    A=\tanh(\mu\pi/2)+\coth(\mu\pi/2)\approx2, \hspace{2em} B=\tanh(\mu\pi/2)-\coth(\mu\pi/2)\approx-0.0159
\end{equation}
The power spectrum for the $\times$ mode thus reads,
\begin{equation}\label{7.50}
    P_{\times}(k)\propto k^3(A+B\cos(2\mu\ln(-k\eta/2)-2\phi_{\mu,0}))
\end{equation}
A plot of the ratio of squared modulus of the two tensor polarizations at superhorizon scales is shown in Fig. (\ref{fig:16}).
We can see that as the modes grow beyond the horizon further and further, the ratio is decaying rapidly. The + mode is nearly a constant on superhorizon scales. But the $\times$ mode is decaying rapidly and hence may not be relevant from the observational point of view. As it is shown in Fig. (\ref{fig:16}), the two polarizations can be distinguished by their behaviour on super horizon scales. Also, the power spectrum for the $\times$ mode has a strong scale dependence in contrast to the past work in antisymmetric tensor field inflation \textcolor{blue}{\cite{Aashish:2020mlw}}. The general scalar field inflation models predict the two tensor polarizations to behave identically while giving a nearly scale invariant power spectrum  \textcolor{blue}{\cite{Riotto:2002yw}}.
Chiral gravitational waves are generally expected in models with an underlying parity violation \textcolor{blue}{\cite{Adshead:2013qp,Bartolo:2018elp,Takahashi:2009wc}}. The parity violation leading to a specific handedness causes the asymmetry in the two gravitational wave polarizations. In our case, we could not sight any parity violating terms in the tensor field action. The asymmetry that occurs in our case has to be investigated more and will be addressed in the following works.


\section{Conclusion} \label{sec:8}
We analyzed the implications of working with a new choice of the background structure for an antisymmetric tensor field driven inflation model. Unlike past results \textcolor{blue}{\cite{Aashish:2018lhv}} slow roll inflation with enough number of e-folds is supported without the need for nonminimal coupling to gravity. We studied the perturbations, initially only in the driving field $B_{\mu\nu}$ and then in the tensor sector of the metric $g_{\mu\nu}$. The model (\ref{2.5}) is free of ghost instabilities. Further, we looked at the primordial gravitational waves that generate from the tensor perturbations in the metric. The velocity of these waves was found to be $c/3$, which is one third of the speed of light in vacuum and differs from the recent GW data which constrains the GW speed to around $c$. Although, the GW speed constraints are only valid for astrophysical sources, and are not technically applicable to primordial gravitational waves. Nevertheless, we showed that the GW speed constraint can be satisfied by adding a nonminimal coupling term $R B_{\mu\nu} B^{\mu\nu}$ in the action (\ref{2.5}). Analyzing the evolution of these tensor modes, we could see that they have an oscillatory behaviour in the subhorizon limit. But, on superhorizon scales these modes were nearly frozen in time, yielding a nearly scale invariant power spectrum.
\\
\\
Motivated by the requirement of graceful exit to reheating phase, we considered a new potential which is quartic in $B_{\mu\nu}$, given by Eq. (\ref{7.1}). With this potential, inflation can only be supported with the help of a non-minimal coupling. Further, the study of primordial gravitational waves leads to an interesting observation that the $+, \times$ polarizations evolve differently in the superhorizon limit. The $+$ polarization remains almost frozen in time and yields a nearly scale invariant power spectrum. But the $\times$ polarization possesses an oscillatory behaviour that decays in time. The resulting power spectrum has a strong dependence on scales, causing the $\times$ mode to decay rapidly compared to + mode. We don't yet have a physical explanation as to why the two tensor modes evolve differently, however. 
\\
\\
An obvious next step in this analysis is the study of vector and scalar perturbations, which will be dealt with in upcoming works. Moreover, phenomenological studies need to be performed including a general background structure of $B_{\mu\nu}$. 
In our model, the isotropy and homogeneity is not inherent, but imposed through the constraint equations. We aim to explore alternate scenarios in the future where the homogeneity and isotropy is inherent.

\section*{Acknowledgements}
This work was partially funded by DST (Govt. of India), Grant No. SERB/PHY/2021057.

\bibliographystyle{ieeetr}
\bibliography{refs}

\end{document}